\documentclass[a4paper, accepted=2023-09-27]{quantumarticle}
\pdfoutput=1

\usepackage{amsmath}
\usepackage[normalem]{ulem}
\usepackage[dvipsnames]{xcolor}
\usepackage{amsthm}
\usepackage{amssymb}
\usepackage{bm}
\usepackage{physics}
\usepackage{dsfont}
\usepackage[toc,page]{appendix}
\usepackage{algorithm}
\usepackage{algpseudocode}
\usepackage{mathtools}

\usepackage{graphicx}
\usepackage{dcolumn}
\usepackage{bm}
\usepackage{hyperref}
\usepackage[capitalise]{cleveref}
\usepackage{float}
\newfloat{algorithm}{t}{lop}
\usepackage{caption}
\usepackage{subcaption}
\usepackage{color,soul}

\begin{document}

\title{Unbiasing time-dependent Variational Monte Carlo by projected quantum evolution}

\author[1,2]{Alessandro Sinibaldi}
\author[1,2]{Clemens Giuliani}
\author[1,2]{Giuseppe Carleo}
\author[1,2,3]{Filippo Vicentini}
\affil[1]{Institute of Physics, \'{E}cole Polytechnique F\'{e}d\'{e}rale de Lausanne (EPFL), CH-1015 Lausanne, Switzerland}
\affil[2]{Center for Quantum Science and Engineering, \'{E}cole Polytechnique F\'{e}d\'{e}rale de Lausanne (EPFL), CH-1015 Lausanne, Switzerland}
\affil[3]{CPHT, CNRS, \'{E}cole polytechnique, Institut Polytechnique de Paris, 91120 Palaiseau, France}

\begin{abstract}
\noindent
We analyze the accuracy and sample complexity of variational Monte Carlo approaches to simulate the dynamics of many-body quantum systems classically. 
By systematically studying the relevant stochastic estimators, we are able to:
(i) prove that the most used scheme, the time-dependent Variational Monte Carlo (tVMC), is affected by a systematic statistical bias or exponential sample complexity when the wave function contains some (possibly approximate) zeros, an important case for fermionic systems and quantum information protocols; 
(ii) show that a different scheme based on the solution of an optimization problem at each time step is free from such problems;
(iii) improve the sample complexity of this latter approach by several orders of magnitude with respect to previous proofs of concept.
Finally, we apply our advancements to study the high-entanglement phase in a protocol of non-Clifford unitary dynamics with local random measurements in 2D, first benchmarking on small spin lattices and then extending to large systems. 
\end{abstract}

\maketitle

\section{Introduction}
\label{sec:intro}

The boundaries of current computational paradigms shape the scope of questions that can be investigated in many-body quantum systems.
Problems such as the dynamics of high-dimensional interacting systems~\cite{georgescu2014quantum}, dissipative phase transitions out of equilibrium~\cite{minganti2021continuous}, the simulation of digital quantum circuits~\cite{nielsen_chuang_2010}, or quantum information protocols~\cite{skinner2019measurement,li2019measurement,tang2020measurement,turkeshi2020measurement,turkeshi2021measurement,lavasani2021topological,lunt2021measurement,liu2022measurement,turkeshi2022entanglement,sierant2022measurement,hoke2023quantum} on states with volume-law entanglement all suffer from a lack of efficient computational methods.  
A class of powerful numerical techniques to treat such problems are variational methods, which rely on an efficient parametrization of the quantum state and stochastic optimization of its parameters.
Compared to more established Tensor Network (TN)~\cite{white1992density,orus2014practical} or Quantum Monte Carlo (QMC)~\cite{ceperley1986quantum,foulkes2001quantum} algorithms, variational approaches coupled with Monte Carlo sampling can  simulate high-dimensional or unstructured systems while not suffering from the sign-problem~\cite{troyer2005computational}, making them ideal candidates to target molecules~\cite{Choo2020Fermions,Pfau2020PRR} and fermionic~\cite{NomuraPRBFermionic,Stokes2020PRB,Nys22Fermions} or frustrated matter~\cite{choo2019two}.
However, the optimization problem arising in variational calculations is generally non-convex, making it hard to give general convergence guarantees.
While state-of-the-art results for the calculation of ground states~\cite{choo2019two,sharir2020deep,varbench} have been obtained with Variational Monte Carlo (VMC)~\cite{mcmillan1965ground}, variational simulations of dynamics have yet to improve over existing approaches systematically. 

Techniques for the variational time evolution rely on so-called \textit{variational principles} to recast Schrödinger's differential equation for the wave function onto non-linear differential equations for the parameters~\cite{Yuan2019TDVP}.
These latter equations can be integrated with an explicit scheme,  the \emph{time-dependent Variational Monte Carlo} or tVMC~\cite{Carleo2012TDVP,Carleo2017tVMC}, or with an implicit method by solving an optimization problem at every time-step~\cite{jonsson2018neural,medvidovic2021classical,Barison2021pVQD,donatella2022dynamics,gutierrez2022real}.
The first approach has been applied to many systems~\cite{schmitt2020quantum,verdel2021variational,schmitt2022quantum}, but it struggled to significantly improve upon benchmark methods due to several poorly-understood challenges in the numerical integration.
The second method is conceptually more powerful than tVMC, but it has yet to be applied to realistic systems due to an unexpectedly large computational overhead~\cite{donatella2022dynamics}. 

\begin{figure*}[ht!]
    \centering
    \includegraphics[scale=0.73]{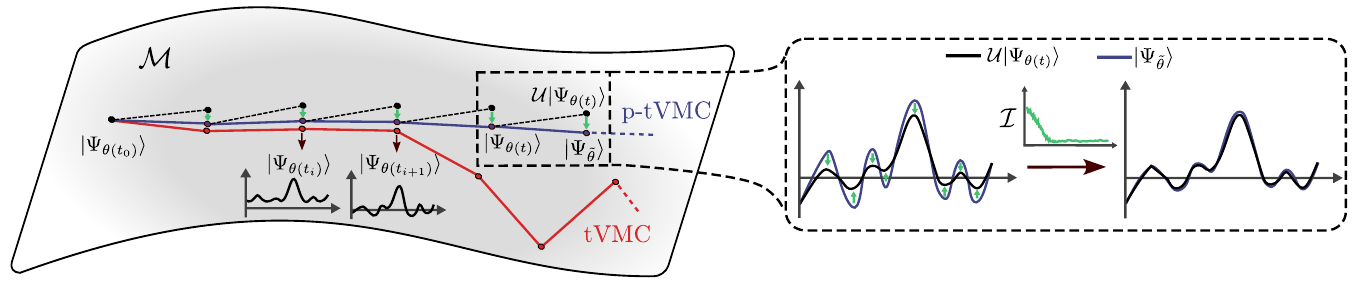}
    \caption{Sketch of the failure of the tVMC when the state features zeros (red) and of the dynamics generated by the p-tVMC algorithm (blue). When a state with zeros (or near zeros) in the wave function is encountered during the tVMC evolution, such as $\ket*{\Psi_{\theta(t_{i+1})}}$, the variational dynamics starts to detach from the exact solution due to a bias (or an vanishing signal to noise ratio). In p-tVMC, the optimization problem of projecting the exactly evolved state $\mathcal{U} \ket*{\Psi_{\theta(t)}}$ onto the variational manifold $\mathcal{M}$ of the ansatz $\ket{\Psi_{\tilde{\theta}}}$ is solved at each time-step. This is achieved by minimizing a distance in the Hilbert space, which is the infidelity $\mathcal{I}$ (as shown in the right panel).}
    \label{fig:artistic_figure}
\end{figure*}

In this manuscript, we systematically analyze the accuracy and efficiency of stochastic variational methods to tackle dynamical problems. 
First, we formalize the origin of the \textit{numerical challenges affecting tVMC} by proving that they arise from the Monte Carlo sampling, which may hide a bias or an exponential cost when the wave function contains zeros, as is the case for many physically-relevant problems.
Then, we prove that the high overhead of the implicit integration arises from poor scaling of the Monte Carlo sampling and we derive a new scheme that lowers the computational cost by several orders of magnitude.
We call this scheme \emph{projected} tVMC (p-tVMC).
Finally, we apply the p-tVMC to simulate the dynamics of a quantum system undergoing unitary evolution interspersed with random measurements, which is a paradigmatic model for entanglement phase transitions~\cite{skinner2019measurement,li2019measurement,tang2020measurement,turkeshi2020measurement,turkeshi2021measurement,lavasani2021topological,lunt2021measurement,liu2022measurement,turkeshi2022entanglement,sierant2022measurement,hoke2023quantum}.
Established methods can access 1D systems~\cite{skinner2019measurement,li2019measurement,tang2020measurement,turkeshi2021measurement,turkeshi2022entanglement,hoke2023quantum} or higher dimensional Clifford dynamics~\cite{turkeshi2020measurement,lavasani2021topological,lunt2021measurement,sierant2022measurement}, but questions remain on the nature of this transition in the case of non-Clifford 2D dynamics. 
As a proof of concept, we investigate this regime which is intractable to TNs due to the rapid entanglement growth and the higher dimensionality but also to tVMC due to the projective measurements enforcing a large number of zeros in the wave function.

\section{Numerical challenges in time-dependent Variational Monte Carlo}
\label{sec:challenges}
We consider a quantum many-body system whose Hilbert space is spanned by the basis states $\{\ket{\sigma}\}$, where $\sigma$ is a set of quantum numbers that is treated as discrete. 
The state of the system $\ket{\Psi}$ can be efficiently approximated by a variational ansatz $\ket{\Psi_{\theta}}$ whose wave function $\Psi_{\theta}(\sigma) \equiv \bra{\sigma} \ket{\Psi_{\theta}}$ is completely specified by a set of $P$ parameters $\theta = (\theta_1, \ldots, \theta_P)$, thus we have: 
\begin{equation}\label{eq:variational_state}
    \ket{\Psi} \approx \ket{\Psi_{\theta}} = \sum_{\sigma} \Psi_{\theta}(\sigma) \ket{\sigma}.
\end{equation}

We consider computationally tractable ansätze, meaning that $P$ is polynomially large in the system size and $\Psi_{\theta}(\sigma)$ can be sampled and queried efficiently~\cite{best2010simulating}.

Within this framework, variational dynamics can be encoded onto time-dependent parameters $\theta(t)$ such that $\ket*{\Psi_{\theta(t)}}$ approximates the physical dynamics. 
In what follows, we focus on the unitary evolution of a time-independent Hamiltonian $\mathcal{H}$, on complex $\theta$ and holomorphic $\Psi_\theta(\sigma)$. 
However, the discussion is general and also applies to non-Hermitian PT-symmetric Hamiltonians~\cite{moiseyev2011non}, imaginary time evolution~\cite{foulkes2001quantum}, or open quantum systems obeying the Lindblad Master Equation~\cite{raimond2006exploring} and can be extended to the non-holomorphic case. 

The McLachlan's variational principle~\cite{mclachlan1964variational} recasts the Schrödinger's equation $\frac{d\ket{\Psi_{\theta}}}{dt} = -i\mathcal{H}\ket{\Psi_{\theta}}$ at every time onto the optimization problem:
\begin{equation}\label{eq:FS_distance}
    \min_{\dot{\theta}}\mathcal{D}( | \Psi_{\theta(t) + \delta t\dot{\theta}(t)} \rangle, e^{-i \mathcal{H} \delta t}|\Psi_{\theta(t)} \rangle),
\end{equation}
where $\mathcal{D}$ is the Fubini-Study metric and $\delta t$ is a small time-step. 
By keeping only the leading terms in $\delta t$ in \cref{eq:FS_distance}, it is possible to derive the following set of explicit equations of motion for $\theta(t)$:
\begin{equation}\label{eq:TDVP-equations}
       \dot{\theta}_{k}(t) = -i\sum_{k'} ( S^{-1} )_{k k'}F_{k'}.
\end{equation}

$F_k$ are the variational forces and $S_{k k'}$ is the Quantum Geometric Tensor~\cite{stokes2020quantum,park2020geometry,hackl2020geometry}. 
These two quantities are defined as: 
\begin{align}\label{eq:F}
F_k &\textrm{=} \frac{\bra{{\partial_{\theta_k} \Psi_{\theta}}} \mathcal{H} \ket{\Psi_{\theta}}}{\bra{\Psi_{\theta}}\ket{\Psi_{\theta}}}  \text{\scalebox{0.95}[1.0]{$-$}} \frac{\bra{\partial_{\theta_k} \Psi_{\theta}}\ket{ \Psi_{\theta}}}{\bra{\Psi_{\theta}} \ket{\Psi_{\theta}}} \frac{\bra{\Psi_{\theta}} \mathcal{H} \ket{\Psi_{\theta}}}{\bra{\Psi_{\theta}} \ket{\Psi_{\theta}}}, \\
\label{eq:S}
\hspace{-1cm}S_{k k'} &\textrm{=} \frac{\bra{\partial_{\theta_{k} }\Psi_{\theta}}\ket{\partial_{\theta_{k'}} \Psi_{\theta}}}{\bra{\Psi_{\theta}} \ket{\Psi_{\theta}}}  \text{\scalebox{0.95}[1.0]{$-$}} \frac{\bra{\partial_{\theta_k} \Psi_{\theta}} \ket{\Psi_{\theta}}}{\bra{\Psi_{\theta}} \ket{\Psi_{\theta}}} \frac{\bra{\Psi_{\theta}}\ket*{\partial_{\theta_{k'}} \Psi_{\theta}}}{\bra{\Psi_{\theta}} \ket{\Psi_{\theta}}}, 
\end{align}
where we use the notation $\ket{\Psi_{\theta}} \equiv |\Psi_{\theta(t)}\rangle$. 
The equations of motion \cref{eq:TDVP-equations} are integrated with numerical schemes like Euler or higher-order Runge-Kutta.

The quantities in \cref{eq:F,eq:S} are efficiently computed by estimating the expectation values as statistical averages over the Born distribution $\Pi(\sigma) = |\Psi_{\theta}(\sigma)|^2/\bra{\Psi_{\theta}} \ket{\Psi_{\theta}}$ with Monte Carlo sampling, following the scheme known as \emph{time-dependent Variational Monte Carlo} (tVMC)~\cite{Carleo2012TDVP,Carleo2017tVMC}. 
The resulting stochastic expressions are given by: 
\begin{align}\label{eq:MC_F}
    F_k^{\text{MC}} &= \mathbb{E}_{\Pi} [ O^*_k(\sigma) ( E_{\mathrm{loc}}(\sigma) - \mathbb{E}_{\Pi} [E_{\mathrm{loc}}(\sigma)] ) ] \\
    \label{eq:MC_S}
    S_{k k^\prime}^{\text{MC}} &= \mathbb{E}_{\Pi} [O^*_k(\sigma) (O_{k^\prime} (\sigma) - \mathbb{E}_{\Pi}[O_{k^\prime} (\sigma)] ) ]. 
\end{align} 

In the previous relations, the quantity $E_{\mathrm{loc}}(\sigma) = \sum_{\sigma^{\prime}} \mathcal{H}_{\sigma, \sigma^\prime} \Psi_{\theta}(\sigma^{\prime})/\Psi_{\theta}(\sigma)$ is the local energy and  $O_k(\sigma)=\partial_{\theta_k} \log \Psi_\theta(\sigma)$ are the log-derivatives of the variational state.

However, we remark that if the wave function and its derivatives have non-identical support, namely there exist configurations $\sigma$ for which $\Psi_{\theta}(\sigma)$ is zero but $\partial_{\theta_k} \Psi_{\theta}(\sigma)$ is not, the stochastic estimate $F_k^{\text{MC}}$ and $S_{k k^\prime}^{\text{MC}}$ may differ from the \textit{exact} quantities in \cref{eq:F,eq:S} by \textit{bias terms} as:
\begin{align}
    \label{eq:bias_in_F}
    &F_{k} = \underbrace{\sum_{\sigma : |\Psi_{\theta}(\sigma)| = 0}  \frac{\bra{\partial_{\theta_k} \Psi_{\theta}} \ket{\sigma} \langle \sigma | \mathcal{H}\ket{\Psi_{\theta}}}  {\bra{\Psi_{\theta}} \ket{\Psi_{\theta}}}}_{\textstyle \text{bias} \, b_F} + F_{k}^{\text{MC}}, \\
    \label{eq:bias_in_S}
    &S_{k k'} = \underbrace{\sum_{\sigma : |\Psi_{\theta}(\sigma)| = 0}  \frac{\bra{\partial_{\theta_k} \Psi_{\theta}} \ket{\sigma} \langle \sigma | \partial_{\theta_{k^\prime}} \Psi_{\theta}\rangle}  {\bra{\Psi_{\theta}} \ket{\Psi_{\theta}}}}_{\textstyle \text{bias} \, b_S} + S_{k k^\prime}^{\text{MC}}.
\end{align}

The previous relations show that $F_k^{\text{MC}}$ and $S_{k k^\prime}^{\text{MC}}$ are biased if $\bra{\sigma} \ket{\Psi_{\theta}}$ vanishes on some configurations while $\bra{\sigma} \ket*{\partial_{\theta_k}\Psi_{\theta}}$ do not, leading to a mismatch between the tVMC and ideal variational dynamics.

\begin{figure}[ht!]
    \centering
    \includegraphics[scale = 0.13]{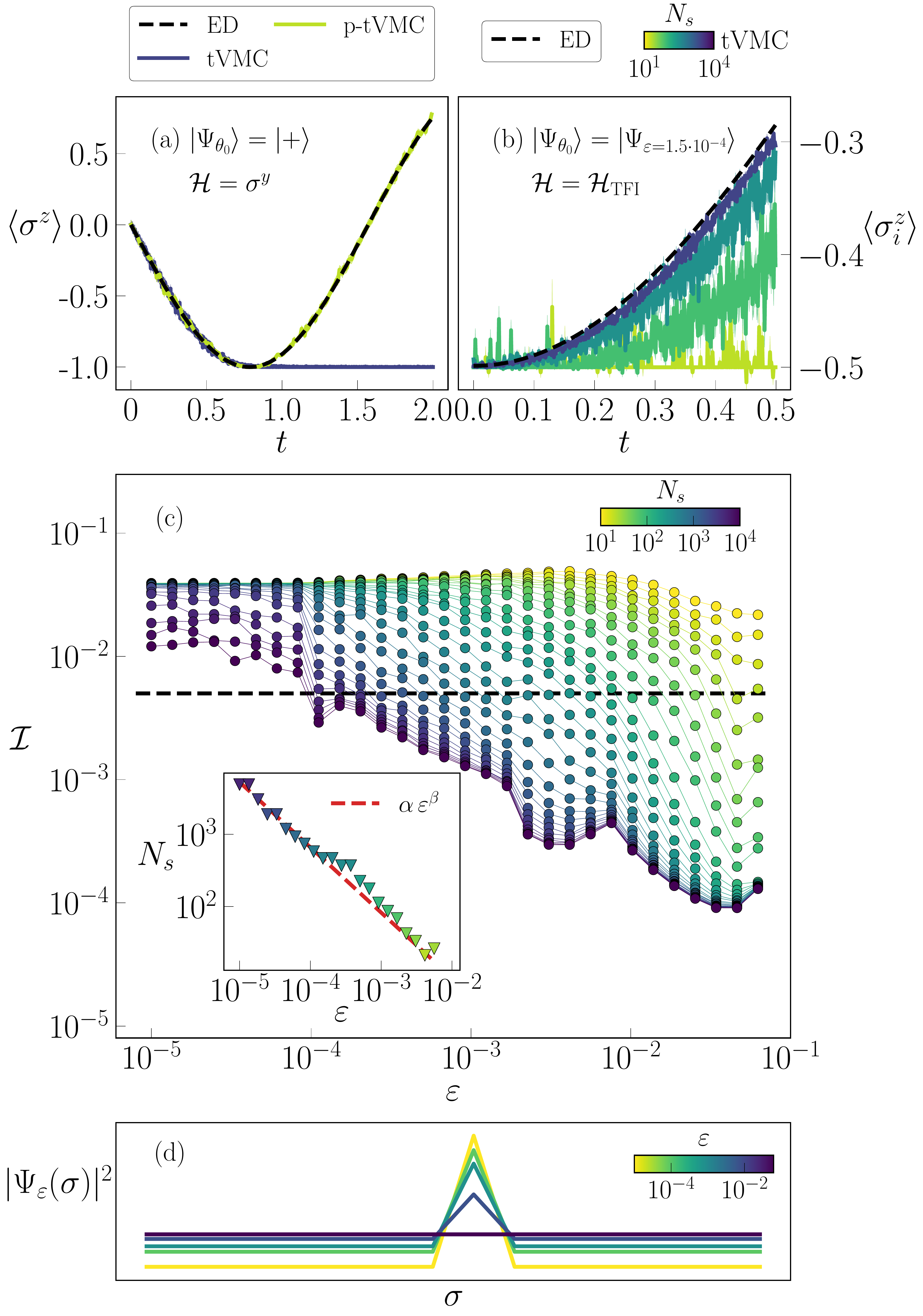}
    \caption{\textbf{(a-b)} Dynamics of the $z$-magnetization $\langle \sigma_i^z \rangle$ for: (a) the $N=1$ spin $1/2$ toy-model, where the initial state $\ket{+}$ is rotated with $\sigma^y$, which is simulated by Exact Diagonalization (ED), tVMC and p-tVMC; (b) a system of $N=4$ spins $1/2$, initialized in $\ket{\Psi_{\varepsilon}}$ with $\varepsilon = 1.5 \cdot 10^{-4}$ and evolved via $\mathcal{H}_{\text{TFI}}$ ($J = h = 1$), which is simulated by ED and tVMC with increasing number of samples $N_s$ (see colorbar). 
    For the tVMC the time-step $\delta t=10^{-3}$ has been used, while for the p-tVMC $\delta t=10^{-2}$.
    \textbf{(c)} Infidelity $\mathcal{I}$ among states evolved with tVMC and with ED after a time $t_f$ starting from $\ket{\Psi_{\varepsilon}}$ with increasing $N_s$ (see colorbar) and different $\varepsilon$. 
    The inset shows how the minimal $N_s$ to reach $\mathcal{I} = 5 \cdot 10^{-3}$ (indicated by a dashed line in (c)) scales with $\varepsilon$ (markers) and a power-law fit (red line). 
    In (a) for tVMC the ansatz parametrizing the wave function amplitudes is used, while in (b-c) a Restricted Boltzmann Machine (RBM)~\cite{carleo2017Science} with $\alpha = 1$. \textbf{(d)} Illustration of the Born distribution for the peaked states $\ket{\Psi_{\varepsilon}}$. }
    \label{fig:scales_with_epsilon}
\end{figure}

This condition may arise from the variational encoding of several physically relevant states, such as basis states $\ket*{\sigma}$, anti-symmetric wave functions (e.g. Slater, Neural Backflow~\cite{diluo_neuralbackflow}, \dots) or states generated by digital quantum circuits.
Another relevant class of affected states are those that underwent projective measurements, which are commonly found in trajectory unravelings of the Lindblad Master Equation~\cite{molmer1993monte,wiseman_milburn_2009,Minganti2022PRABetadine} or in quantum information measurement protocols~\cite{skinner2019measurement,li2019measurement,tang2020measurement,turkeshi2020measurement,turkeshi2021measurement,lavasani2021topological,lunt2021measurement,liu2022measurement,turkeshi2022entanglement,sierant2022measurement,hoke2023quantum}. 
We remark that for continuous systems (such as a particle in free space), the bias may only emerge from zeros in the bulk, since at infinity the wave function and its derivative must vanish.
A pictorial representation of the breakdown of tVMC is shown in \cref{fig:artistic_figure}. 

In realistic calculations, the variational wave function is often arbitrarily close to zero without encoding nodes exactly.
In such cases, the biases $b_F$ and $b_S$ are zero. Still, we find that the variances of $F^{\text{MC}}$ and $S^{\text{MC}}$ grow such that an exponential number of samples is required to resolve those quantities with finite accuracy.
This phenomenon can be revealed by the signal-to-noise ratios (SNRs) of $F^{\text{MC}}$ and $S^{\text{MC}}$ approaching zero. 
The SNR of a function $f$ of random variable $\sigma$ with distribution $\Pi$ is: 
\begin{equation}
    \text{SNR}_\Pi[f] = \sqrt{\frac{|\mathbb{E}_{\Pi}[f(\sigma)]|^2}{\text{Var}_{\Pi}[f(\sigma)]}}.
\end{equation}

In practical calculations, 
to ensure an accurate estimation with a finite number of samples $N_s$ we require $\text{SNR}_{\Pi}[f] \sqrt{N_s}\gtrapprox  1$. 
This inequality guarantees that the effective signal (mean value) is larger than the statistical fluctuations and thus it can be resolved. 

In the following, we first discuss a minimal example where finite biases emerge, and we then consider a more realistic case where there are no biases but for which we show that the SNRs go to zero. 

\subsection{Paradigmatic examples}
We analyze a toy model where the biases are non-zero, and they break the tVMC dynamics.
The system is made by $N=1$ spin $1/2$ where the state is parameterized by the ansatz $\ket*{\Psi_{(\alpha, \beta)}} = \alpha \ket{\downarrow} + \beta \ket{\uparrow}$.
For $ \ket{\Psi_{\theta}} = | \Psi_{(1, 0)} \rangle = \ket{\downarrow}$, such that $\bra*{\uparrow} \ket*{\Psi_{\theta} } = 0$, and evolution generated by Hamiltonian $\mathcal{H} = \sigma^y$, we have that $b_F$ and $b_S$ in \cref{eq:bias_in_F,eq:bias_in_S} are finite and the stochastic estimates differ from the exact values as:
\begin{align}
F &= \begin{pmatrix}
0 \\
i  \\
\end{pmatrix}, 
&
S &= \begin{pmatrix}
0 & 0 \\
0 & 1 \\
\end{pmatrix}, \\
F^{\text{MC}} &= \begin{pmatrix}
0 \\
0 \\
\end{pmatrix}, 
&
S^{\text{MC}} &= \begin{pmatrix}
0 & 0 \\
0 & 0 \\
\end{pmatrix}, 
\end{align}
(see \cref{sec:singlespin} for the full calculation).

In \cref{fig:scales_with_epsilon}(a) we show a simulation of an initial state $\ket{+}$ which is rotated by the Hamiltonian $\mathcal{H} = \sigma^y$. 
At $t=\pi / 4$ the state becomes $\ket{\Psi_{\theta}} = \ket{\downarrow}$ and the tVMC evolution is stuck, as $F^{\text{MC}}$ and $S^{\text{MC}}$ vanish. 
Similar considerations hold for more than 1 particle, and in \cref{sec:twospins} we discuss an example for the GHZ state of $N=2$ spins.

We now analyze a more realistic case where the variational state does not exactly encode zeros. 
In particular, we consider a system of $N$ spins in the state $\ket{\Psi_\varepsilon}$, which is peaked on a single configuration $\ket{\sigma_0}$ and has a constant small amplitude $\sqrt{\varepsilon}$ for all the other basis states (see \cref{fig:scales_with_epsilon}(d)), namely: 
\begin{equation}
    \Psi_{\varepsilon}(\sigma) = 
    \begin{cases}
    \sqrt{\varepsilon}, &\text{if $\sigma \neq \sigma_0$},  \\
    \sqrt{1 - (2^N - 1)\varepsilon}, &\text{if $\sigma = \sigma_0$}, 
    \end{cases}
\end{equation}
where $0 < \varepsilon < 1/(2^N - 1)$. 
We remark that for small $\varepsilon$ this ansatz approximates a Hartree-Fock state, which commonly arises from quantum-chemistry Hamiltonians.
For $\varepsilon \neq 0$ the biases $b_{F/S}$ are zero, but for $\varepsilon\approx 0$ the leading terms of the SNRs for $F^{\text{MC}}$ and $S^{\text{MC}}$ are respectively (see \cref{sec:exploding_variance} for the full calculation): 
\begin{equation}
    \text{SNR}_{\Pi}[F^{\text{MC}}] \propto \sqrt{\varepsilon}, \quad
    \text{SNR}_{\Pi}[S^{\text{MC}}] \propto \sqrt{\varepsilon} .
\end{equation}

Intuitively, this suggests that the more peaked the state is, the more samples will be needed to accurately estimate those quantities, in particular $N_s \propto \varepsilon^{-1}$.
As normalization of the state imposes $\varepsilon\propto2^{-N}$, the number of samples necessary to correctly compute the quantum geometric tensor and the variational forces will diverge as $N_s \propto 2^{N}$, eliminating the advantage of stochastic sampling and rendering tVMC computationally ineffective.

We consolidate this argument with a numerical experiment involving a commonly adopted setup for quantum dynamics.
We evolve with tVMC the state $\ket{\Psi_\varepsilon}$ for $t \in [0, t_f]$ according to the Transverse Field Ising (TFI) Hamiltonian,
\begin{equation}\label{eq:TFIM}
    \mathcal{H}_{\mathrm{TFI}}=-J \sum_{\langle i, j\rangle} \sigma_i^z \sigma_j^z - h \sum_i \sigma_i^x,
\end{equation}
where $\langle i, j \rangle$ denotes nearest neighbors in a lattice with periodic boundary conditions.
In \cref{fig:scales_with_epsilon}(b) we show some evolutions obtained with an increasing number of Monte Carlo samples $N_s$ for a fixed $\varepsilon$, demonstrating that the dynamics is correctly reconstructed only at large values of $N_s$. 
The scaling of $N_s$ with the system size is studied in \cref{fig:scales_with_epsilon}(c), where we report the final infidelity\footnote{The infidelity between two arbitrary states $\ket{\psi}$ and $\ket{\phi}$ is defined as
$\mathcal{I}(\psi, \phi)=1-\abs{\braket{\psi}{\phi}}^2/\bra{\psi} \ket{\psi} \bra{\phi} \ket{\phi}$.} of the state obtained with tVMC with respect to the exact solution for different $\varepsilon$ and $N_s$. 
We remark that the accuracy of the variational simulation improves when $\varepsilon$ or $N_s$ are increased, as the statistical fluctuations in the estimated quantities are suppressed.
The inset highlights a power-law relation between the $N_s$ necessary to reconstruct the dynamics accurately and $\varepsilon$,  proving that $N_s \sim 2^{N}$.

\subsection{Overview}
In this first section, we have shown that the tVMC method can be either biased or require an exponential number of samples when the wave function is exactly or approximately zero. 
This highlights the necessity of an efficient alternative method to tVMC for variational time evolution. 

We stress that while our considerations on stochastic estimators arose in the context of tVMC, they are also applicable to ground-state calculations using both plain gradient descent or stochastic reconfiguration~\cite{sorella2007weak}, because they rely on the same stochastic estimators.
However, we believe that in such calculations, the additional errors contributed by the biasing or small SNR are mitigated by the iterative optimization scheme, which may avoid the accumulation of errors that instead affects dynamics.
We also remark that Monte Carlo variational methods for open quantum systems~\cite{vicentini2019prl,Hartmann2019PRLDissipative,nagy2019variational,Reh2021} are also possibly affected by the same issues.

Going forward, in \cref{sec:unbiased_F} we propose a modified estimator for the forces $F$ for which the bias and the SNR problems are absent, and therefore it can efficiently estimate the forces when the standard estimator $F^{\text{MC}}$ fails (see \cref{sec:bias_examples,sec:exploding_variance}). 
From our knowledge, this alternative estimator has never been discussed in the literature, and preliminary investigations suggest that it already reduces the computational effort needed to reliably find the ground state of some frustrated or fermionic Hamiltonians.

Unfortunately, we could not find a similarly straightforward modification of the estimator for the quantum geometric tensor $S$. For that reason, the following section presents a completely different scheme that avoids using the tensor.

\section{Projected time-dependent Variational Monte Carlo}
\label{p-tVMC}

\begin{figure*}[ht!]
     \centering
     \begin{subfigure}[b]{0.20\textwidth}
         \centering
         \includegraphics[scale=0.125]{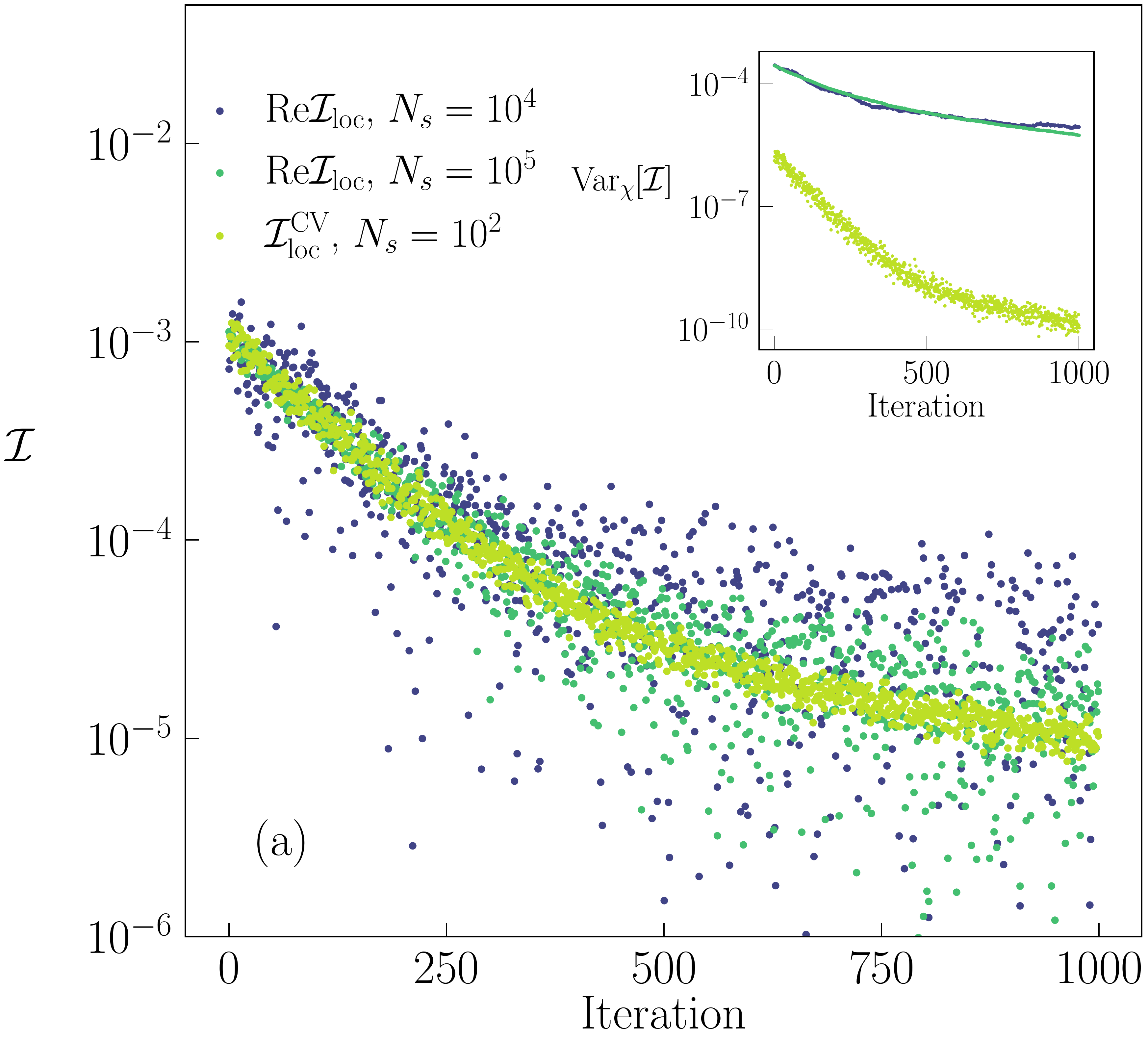}
     \end{subfigure}
     \hfill
     \begin{subfigure}[b]{0.6\textwidth}
         \centering
         \includegraphics[scale=0.125]{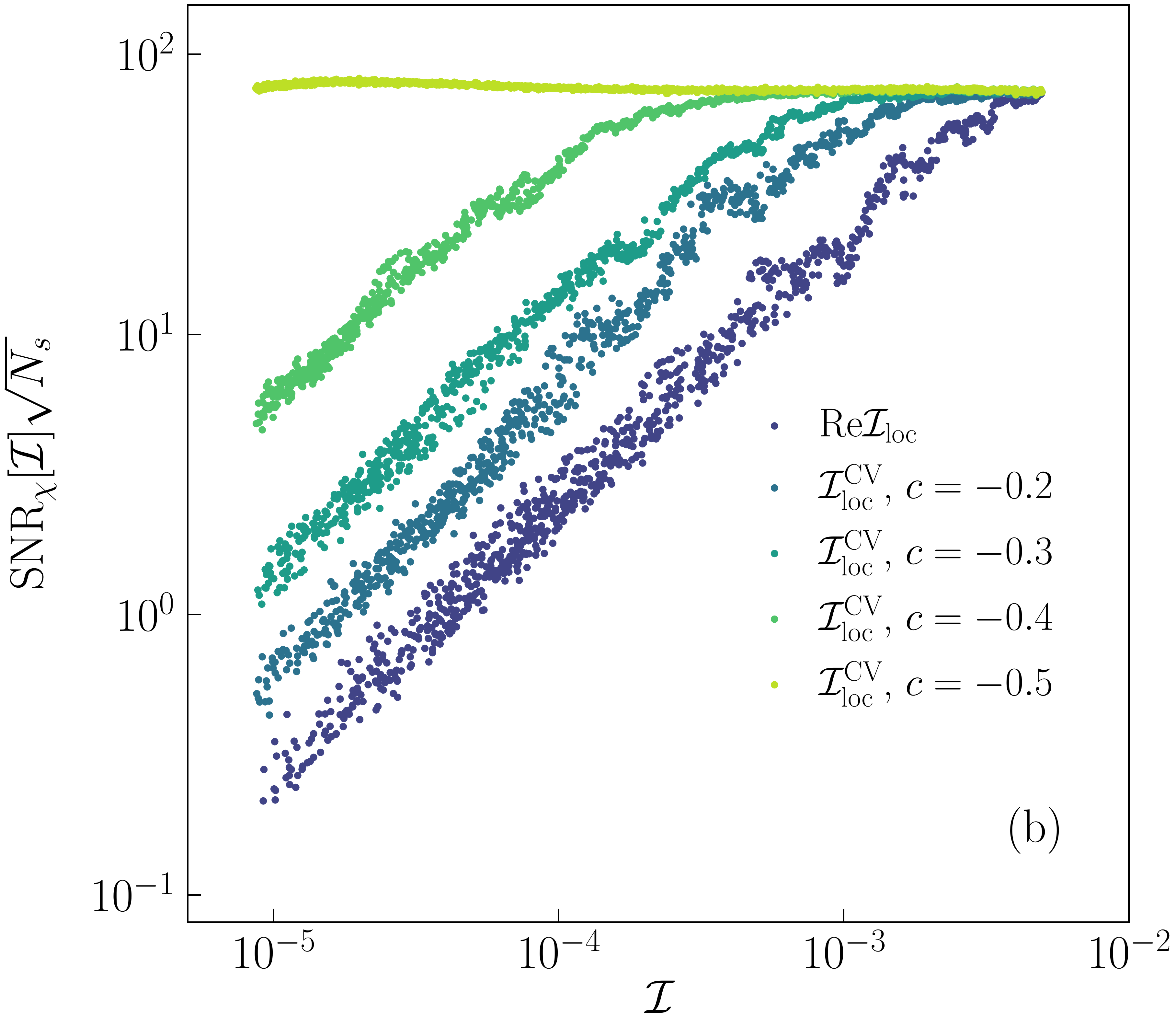}
     \end{subfigure}
    \caption{\textbf{(a)} Learning curves for the infidelity $\mathcal{I}$ using the bare estimator $\Re \mathcal{I}_{\text{loc}}$ and the CV estimator $ \mathcal{I}^{\text{CV}}_{\text{loc}}$. In the inset the corresponding variances, generically indicated as $\text{Var}_{\chi}[\mathcal{I}]$, are shown.
    \textbf{(b)} Rescaled signal to noise ratio of the infidelity $\text{SNR}_{\chi}[\mathcal{I}] \sqrt{N_s}$ as a function of $\mathcal{I}$ for the bare and the CV estimators with different values of $c$.
    We remark that the slope of the curves of the SNR with $c\neq-0.5$ in the limit $\mathcal{I} \rightarrow 0 $ is approximately 1, which is better than what is given by \cref{eq:SNR_bare_estimator}.
    This is because the estimator used in practical calculations is $\Re \mathcal{I}_{\text{loc}}$ instead of simply $\mathcal{I}_{\text{loc}}$ (further discussion in \cref{sec:CV_estimator}). 
    In both (a) and (b) a system of $N=8$ spins $1/2$ is considered
    and the transformation used is $\mathcal{U} = \exp(-i \frac{\delta t}{2} \mathcal{H}_{zz})$, where $\mathcal{H}_{zz} =-J \sum_{\langle i, j\rangle} \sigma_i^z \sigma_j^z$ with $J=1$ and $\delta t = 5 \cdot 10^{-2}$.
    In (b) $N_s = 10^4$ samples have been employed.}
    \label{fig:CV}
\end{figure*}

We consider the general problem of finding the parameters of a variational state $\ket*{\Psi_{\Tilde{\theta}}}$ such that it approximates the state $\mathcal{U}\ket{\Psi_{\theta}}$, where $\theta$ are known and $\mathcal{U}$ is an arbitrary transformation, in terms of a given distance.
Considering the distance to be the infidelity $\mathcal{I}$, this can be expressed as the following optimization problem:
\begin{equation}\label{eq:optimization_problem}
 \min_{\Tilde{\theta}} \mathcal{I}(\ket*{\Psi_{\Tilde{\theta}}},\, \mathcal{U} \ket{\Psi_{\theta}}).
\end{equation}

Other distance choices, such as the L2 metric, have also been discussed in the literature~\cite{gutierrez2022real}.
\cref{eq:optimization_problem} is similar to \cref{eq:FS_distance}, but it can treat arbitrary unitaries and therefore can be used to simulate non-infinitesimal gates in quantum circuits~\cite{jonsson2018neural,medvidovic2021classical} or to perform state preparation.

The solution of \cref{eq:optimization_problem} can be found with iterative gradient-based optimizers such as Stochastic Gradient Descent~\cite{sutskever2013importance}, ADAM~\cite{kingma2017adam}, Natural Gradient~\cite{amari1998natural,sorella2005wave,stokes2020quantum} or similar methods.
Since this approach consists of projecting the exactly evolved state $\mathcal{U} \ket{\Psi_{\theta}}$ onto the manifold of the variational ansatz $\ket{\Psi_{\Tilde{\theta}}}$, we name it \emph{projected time-dependent Variational Monte Carlo} (p-tVMC), and this is pictorially represented in \cref{fig:artistic_figure}. 

The infidelity in \cref{eq:optimization_problem} can be estimated through Monte Carlo sampling as $\mathcal{I}(\Tilde{\theta}) = \mathbb{E}_{\chi}[\mathcal{I}_{\text{loc}}(\sigma, \eta)]$.
Many choices for the sampling distribution $\chi$ and the local estimator $\mathcal{I}_{\text{loc}}$ are possible, but assuming that $\mathcal{U}$ is unitary we can sample
\begin{equation} \label{eq:estimator}
\mathcal{I}_{\text{loc}}(\sigma, \eta) = 1 - \frac{\bra{\sigma} \mathcal{U} \ket{\Psi_{\theta}}}{\bra{\sigma} \ket*{\Psi_{\Tilde{\theta}}}} \frac{\bra{\eta} \mathcal{U}^{\dagger} \ket*{\Psi_{\Tilde{\theta}}}}{\bra{\eta} \ket{\Psi_{\theta}}},
\end{equation}
from the joint Born distribution of the two states $\chi (\sigma, \eta) = |\Psi_{\Tilde{\theta}}(\sigma)|^2 |\Psi_{\theta}(\eta)|^2  / \bra*{\Psi_{\Tilde{\theta}}} \ket*{\Psi_{\Tilde{\theta}}} \bra{\Psi_{\theta}} \ket{\Psi_{\theta}}$. 
For non-unitary $\mathcal{U}$ it is necessary to sample from $\mathcal{U} \ket{\Psi_{\theta}}$ instead of $\ket{\Psi_{\theta}}$~\cite{jonsson2018neural,medvidovic2021classical}.

We remark that estimating the infidelity using \cref{eq:estimator} is efficient if $\mathcal{U}$ is $K$-local~\cite{yan2012klocal}, namely it acts non-trivially on at most $K$ degrees of freedom (spins, qubits, particles, \ldots), where $K$ is polynomially large in the system size. 
When this is not the case, it can be factored in several sub-terms  $\mathcal{U}=\mathcal{U}_1\dots\mathcal{U}_N$ where each term is $K$-local, and \cref{eq:optimization_problem} must be solved for every sub-unitary $\mathcal{U}_i$.
In particular, the unitary propagator of a general Hamiltonian can be decomposed with the Trotter-Suzuki decomposition~\cite{trotter,suzuki} or with other expansions that are unitary up to leading order, such as the Taylor series~\cite{donatella2022dynamics}. 

We now analyze the estimator $\mathcal{I}_{\text{\text{loc}}}$ according to the same approach used in the previous section.
We find that, in the limit of $\mathcal{I} \rightarrow 0$, the $\text{SNR}$ of the estimator scales as
\begin{equation}
\text{SNR}_{\chi}[\mathcal{I}_{\text{loc}}]\propto\sqrt{\mathcal{I}}.
\end{equation}

This means that, as the optimization approaches the optimum of $\mathcal{I} = 0$, the number of samples needed to resolve the infidelity increases as $\mathcal{I}^{-1}$ (see \cref{sec:CV_estimator} for analytical calculation).
This is systematically different from what happens when minimizing the energy, where the SNR remains constant when close to the solution, and a constant number of samples can be used to achieve arbitrarily high precision.

To recover this behaviour in the case of the infidelity optimization we propose a new estimator based upon the \textit{Control Variates} (CV) technique~\cite{rubinstein2016simulation}:
\vspace{0.1cm}
\begin{equation}\label{eq:CV_estimator}
    \mathcal{I}_{\text{loc}}^{\text{CV}}  = \Re \mathcal{I}_{\text{loc}} - c (|1 - \mathcal{I}_{\text{loc}}|^2 - 1).
\vspace{0.1cm}
\end{equation}
where $c \in \mathbb{R}$. 
The quantity $c$ can be chosen such that the $\text{Var}_{\chi}[\mathcal{I}_{\text{loc}}^{\text{CV}}]$ attains a minimal value. 
This optimal value of $c$, say $c^*$, depends on the parameters of the ansatz, so it changes during an infidelity optimization. 
However, it is possible to show (see \cref{sec:CV_estimator} for analytical proof) that in the limit $\ket{\Psi_{\Tilde{\theta}}} \rightarrow \mathcal{U} \ket{\Psi_{\theta}}$, $c^*$ is exactly $-1/2$. 
Therefore, we avoid the high cost of estimating $c^*$ at each iteration of an optimization, and directly use the asymptotically ideal estimator \cref{eq:CV_estimator} with $c=-1/2$. 

To further support this approach, we show in \cref{fig:CV}(a) that the CV estimator $\mathcal{I}_{\text{loc}}^{\text{CV}}$ features a variance that is orders of magnitude smaller than the one of the bare estimator $\Re \mathcal{I}_{\text{loc}}$, in such a way that the number of samples needed for the optimization is reduced by almost three orders of magnitude. 
Additionally, the scaling of the variance with $\mathcal{I}$ changes, such that for $\mathcal{I} \rightarrow 0$ we have that:
\begin{equation}
    \text{SNR}_{\chi}[ \mathcal{I}_{\text{loc}}^{\text{CV}}] = O(1),
\end{equation}
meaning that the SNR remains asymptotically constant (see analytical proof in \cref{sec:CV_estimator}). 
Indeed, in \cref{fig:CV}(b) one can see that the SNR of $\Re \mathcal{I}_{\text{loc}}$ goes to zero when $\mathcal{I}$ decreases, while the SNR of the corrected estimator decreases at smaller values of $\mathcal{I}$. 
When $c = -1/2$, $\mathcal{I}_{\text{loc}}^{\text{CV}}$ has a rescaled SNR which is constant and larger than $1$ over the whole range of $\mathcal{I}$ considered, suggesting that our strategy is ideal.

Moreover, the reduced variance of the CV estimator implies that the infidelity gradient computed with it results to be more accurate~\cite{mohamed2020monte}. 
The lower-variance gradient can improve the accuracy of the solution, since its mean value is affected by smaller statistical fluctuations, and can increase the speed of convergence, as it allows for larger learning rates.

This substantial improvement of the sampling cost using the CV estimator \cref{eq:CV_estimator} makes the p-tVMC an efficiently scalable method for simulating large systems, such that it can address system sizes that have not been investigated before within this approach. 
A detailed analysis on the CV infidelity estimator is present in \cref{sec:CV_estimator}, with further extensions in \cref{sec:importance_sampling}.
As shown in \cref{fig:scales_with_epsilon}(a), in \cref{sec:twospins} for the GHZ state and in \cref{sec:adiabatic} for adiabatic evolution, the p-tVMC, since it is not affected by biases or vanishing SNR, can simulate dynamics in cases where tVMC fails or is inefficient.   

\section{Unitary dynamics with random measurements}
\label{sec:example}
In recent years, considerable interest has been devoted to studying entanglement in many-body quantum systems subject to evolution and random local measurements.
This is a paradigmatic model of a quantum system coupled to an external environment acting as a measurement apparatus. Therefore, it is intimately related to the physics of open quantum systems. 
The competition between the unitary evolution's entangling action and the measurements' localizing effect gives rise to a phase transition between \emph{volume-law} and \emph{area-law} entanglement in the steady state of the dynamics. 
The order parameter is the measurement rate.
This phenomenology has been originally investigated in quantum circuits~\cite{skinner2019measurement,li2019measurement,turkeshi2020measurement,lavasani2021topological,lunt2021measurement,liu2022measurement,sierant2022measurement,hoke2023quantum}, and more recently for continuous dynamics~\cite{tang2020measurement,turkeshi2021measurement,turkeshi2022entanglement}. 

To the extent of our knowledge, numerical investigations have focused so far on systems in one-dimension, integrable or evolving via efficiently simulable~\cite{gottesman1998heisenberg} Clifford gates because of algorithmic limitations.
However, several open questions remain on the nature of such transitions in non-integrable 2D systems or non-Clifford circuits, requiring novel computational paradigms. 

In this concluding section, we leverage the p-tVMC to simulate the time evolution generated by the (non-Clifford) 2D TFI model subject to random local measurements. 
This problem cannot be treated efficiently with Tensor Network methods because of the rapid entanglement growth and the exponential cost of exact contractions in 2D.
Moreover, as projective measurements insert exponentially many zeros in the wave function amplitudes, the shortcomings of tVMC discussed in this article emerge, resulting in an exponential cost. 
We consider a 2D spin $1/2$ square lattice with side length $L$, such that the total number of spins is $N=L^2$. 
We evolve the system to time $t_f$, discretized into time-steps of duration $\delta t$. 
At each $t$, two operations are performed on the system: 
\begin{itemize}
\item unitary evolution with $\mathcal{U} = e^{-i  \mathcal{H}_{\text{TFI}}\delta t}$, where $\mathcal{H}_{\text{TFI}}$ is the 2D TFI Hamiltonian of \cref{eq:TFIM}; 
\item a projective measurement of each spin in the $\sigma^z$ basis independently  with probability $p$ (measurement rate).
\end{itemize}
The overall evolution is stochastic and non-unitary. 

As the unitary propagator is not $K$-local, we decompose it with a second-order Trotter scheme as:
\begin{equation}
\vspace{0.1cm}
e^{-i  \mathcal{H}_{\mathrm{TFI}}\delta t}  =  e^{-i \mathcal{H}_{zz}\frac{\delta t} {2}}  e^{-i \mathcal{H}_x \delta t } e^{-i \mathcal{H}_{zz} \frac{\delta t}{2}} + O(\delta t^3),   
\end{equation}
\vspace{0.1cm}
where $\mathcal{H}_{zz} =-J \sum_{\langle i, j\rangle} \sigma_i^z \sigma_j^z$ and $\mathcal{H}_{x} = - h \sum_i \sigma_i^x $. 

\begin{figure}[ht!]
    \centering
    \includegraphics[scale=0.14]{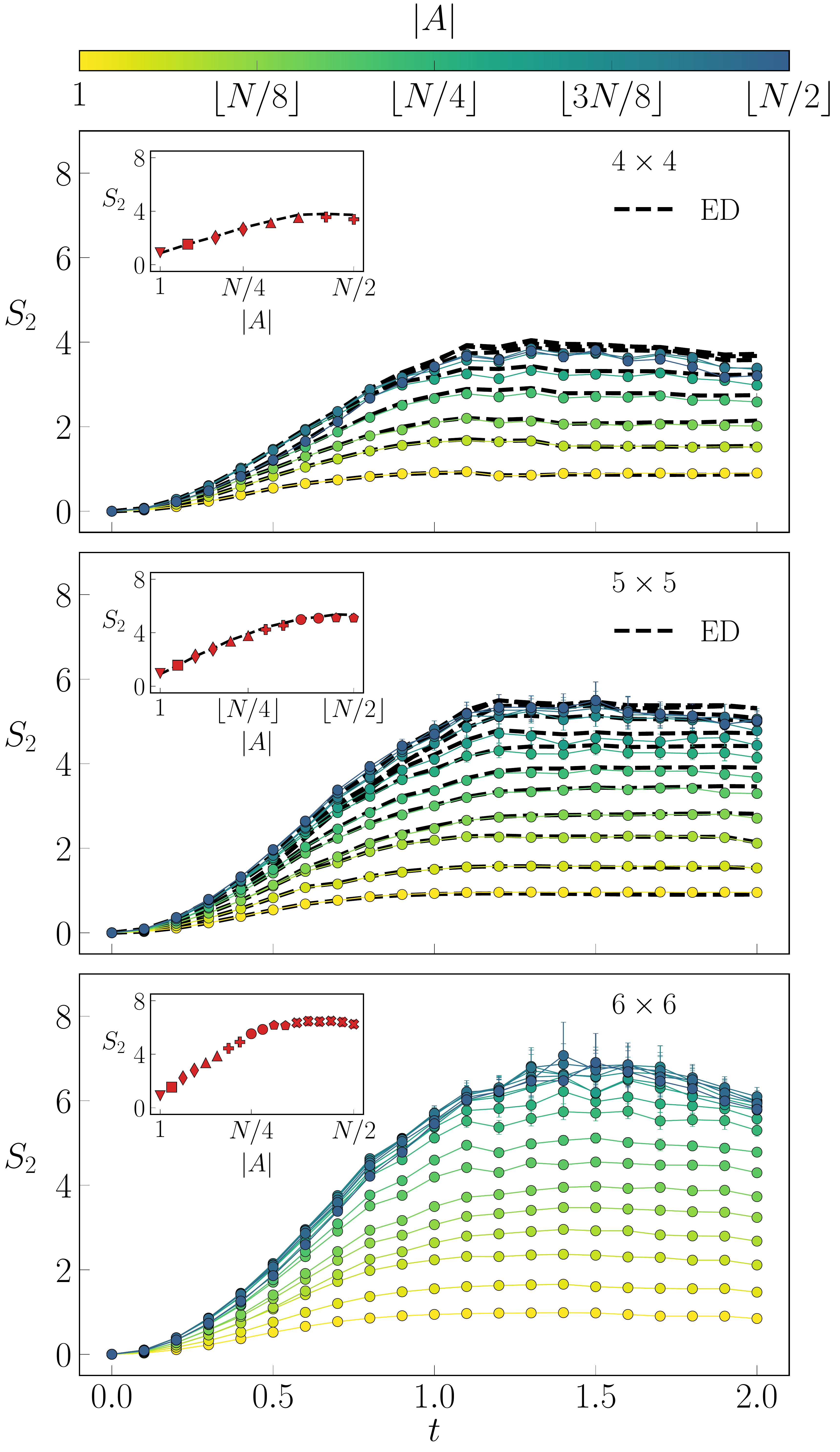}
        \caption{Time evolution of the Rényi-2 entropy $S_2$ simulated with p-tVMC for $\mathcal{H}_{\text{TFI}}$ interspersed with random local measurements along $z$ in 2D $L \times L$ lattices.
        For $L=4$ and $L=5$ we also provide ED benchmark results. 
        Subsystems of increasing size $|A|$ up to the maximum $\lfloor N/2 \rfloor$ are considered, and the corresponding markers are indicated with different colors according to the colorbar. 
        The insets show the scaling of $S_2$ as a function of $|A|$ in the steady-state for the three lattices. $S_2$ for subsystems with equal boundary length is indicated using the same marker. The initial state is $\bigotimes_{i=1}^N \ket{+}_{i}$ and the parameters of $\mathcal{H}_{\text{TFI}}$ are $J=1/2$ and $h=h_c/4$. 
        The measurement rate is $p=0.01$ and the time interval is $\delta t = 0.1$. 
        The results are averaged over $5$ trajectories. 
        The ansatz is an RBM with $\alpha = 4$, endowed with the variational terms to exactly implement the diagonal part of the propagator and the measurements.
        The number of samples is $N_s = 10^4$ for $L = 4$, and $N_s = 2 \cdot 10^4$ for $L=5, 6$. 
        }
        \label{fig:S2}
\end{figure}

In the first stages of this work we employed the forward-backward scheme as done in~\cite{gutierrez2022real,diluo2023}, but then we move to the Trotterization as it allowed to use larger $\delta t$ and was more practical for our calculations.

We use the p-tVMC to apply each unitary obtained from the decomposition. Still, we remark that as $\exp(-i \mathcal{H}_{zz}\delta t /2)$ is diagonal in the chosen $\sigma_z$ basis, its p-tVMC optimization problem can be solved analytically, as shown in \cref{sec:exact_application}.
Instead, the unitary containing $\mathcal{H}_x$ is applied using the p-tVMC, factorizing the propagator into a product of terms, each of which acts on a small subset of the spins. 
In this way, the number of connected elements to compute is not exponentially large with $N$. 
The unitary dynamics that we simulate is a global quench across the critical point of the 2D $\mathcal{H}_{\text{TFI}}$, which is in correspondence of the transverse field $h_c$ such that $h_c/J \approx 3.044$~\cite{elliott_prb_1970_tfi2d,dejongh_prb_1998_tfi2d_dmrg,rieger_epjb_1999_tfi2d_cluster_algo,bloete_pre_2002_tfi2d,albuquerque_alet_prb_2010_tfi2d}. 
In particular, the initial state is the paramagnetic ground state in the limit $h \rightarrow \infty$, given by $\bigotimes_{i=1}^N \ket{+}_{i}$, and this is evolved into the ferromagnetic phase where $h < h_c$. 

The measurement operators are $\mathcal{P}^{\pm}_{i} = (1 \pm \sigma^z_{i})/2$ for $i \in [1, N]$. 
When a spin $i$ is measured, one of the two projectors $\mathcal{P}^{\pm}_{i}$ is applied to the state according to the probabilities $\bra{\Psi}\mathcal{P}^{\pm}_{i}|\Psi\rangle / \bra{\Psi}\Psi\rangle$. 
Measurements can be exactly realized on a variational state $\ket{\Psi_{\theta}}$ by modifying it into a new ansatz $\ket{\Psi_{\phi, \theta}}$ with additional parameters $\phi = (\phi_{1}^{\downarrow}, \phi_{1}^{\uparrow}, \phi_{2}^{\downarrow}, \phi_{2}^{\uparrow}, \ldots)$ defined as: 
\begin{equation}
    \bra{\sigma} \ket{\Psi_{\phi, \theta}} = \prod_{i=1}^{N} (\phi_{i}^{\downarrow} \delta_{\sigma_i, \downarrow}+ \phi_{i}^{\uparrow}\delta_{\sigma_i, \uparrow}) \bra{\sigma }\ket{\Psi_{\theta}}.
\end{equation}

Measuring spin $i$ with outcome $\downarrow$ translates into setting $\phi_{i}^{\uparrow} = 0$ and  $\phi_{i}^{\downarrow} \neq 0$ and vice versa for the other outcome. 
The measurement probabilities are stochastically computed using Monte Carlo sampling. 
The protocol of unitary evolution and random measurements is repeated several times and the final result is obtained by averaging over all these trajectories, as in the Monte Carlo wave function method~\cite{molmer1993monte}. 
To study the entanglement growth of $\ket{\Psi_{\theta}}$, we monitor the Rényi-2 entanglement entropy $S_2(\rho_A) = - \log_2 \Tr \rho_{A}^2$, where $\rho_{A}$ is the reduced density matrix of the state on a subsystem $A$. 
Indeed, the Rényi-2 entropy is a lower bound for the Von Neumann entanglement entropy and it can be estimated via Monte Carlo sampling~\cite{HastingsPRL2010} as:
\begin{equation}\label{Reniy2_stochastic}
    S_2 = -\log_2\Bigg(\mathbb{E}_{\substack{\Pi(\sigma, \eta) \\ \hspace{0.2cm}\Pi(\sigma^\prime, \eta^\prime)}} \bigg[ \frac{\Psi_{\theta}(\sigma^{\prime}, \eta) \Psi_{\theta}(\sigma, \eta^{\prime})}{\Psi_{\theta}(\sigma, \eta) \Psi_{\theta}(\sigma^{\prime}, \eta^{\prime})}\bigg] \Bigg), 
\end{equation}
where $\sigma, \sigma^{\prime} \in A$ and $\eta, \eta^{\prime} \in B$ (complementary of $A$). 
See \cref{sec:Reniy2} for a derivation of \cref{Reniy2_stochastic}. 

\cref{fig:S2} shows the evolution of $S_2(\rho_A)$ for subsystems of size $\abs{A} \in [1, \lfloor N/2 \rfloor]$ in lattices with $L=\{4, 5, 6\}$ and with measurement rate $p=0.01$. 
To assess the quality of the variational simulations, we compare with ED for $L=4, 5$ and we select a feature density for the RBM ansatz employed (which determines its expressivity) that gives a satisfying level of precision.
Given the chosen hyper-parameters, there is an excellent agreement for small subsystem sizes ($\abs{A} \lesssim \lfloor N /4 \rfloor$) and a good agreement for the largest partitions.
The Rényi-2 entanglement entropy is 0 at $t=0$, as expected for the initial product state, and it grows linearly over time, plateauing to a value that is proportional to $|A|$. 
The Page-like curves~\cite{page1993average} in the insets of \cref{fig:S2} suggest that with $p=0.01$ the steady-states belong to a high-entanglement phase. 
However, due to possible finite size effects, whether the scaling of $S_2$ with the subsystem size is linear, witnessing a volume-law, or logarithmic as in critical phases is not obvious.
In any case, a low-entanglement regime with area-law scaling is excluded since, in that situation, the steady-state $S_2$ would not change for subsystems with the same boundary length (indicated with equal markers in the insets). 
Instead, what is observed is that $S_2$ increases with $|A|$ independently of the boundary length, at least far from $\lfloor N/2 \rfloor$ where finite size effects might play a role. 
The proportionality of the entanglement growth rate in the initial times with the boundaries of the partitions is in accordance with the Lieb-Robinson bound~\cite{bravyi2006lieb} valid for local Hamiltonians. 

\section{Conclusions}
\label{sec:conclusions}
In this manuscript, we proved that the standard approach to Monte Carlo variational dynamics, the tVMC, can be limited by a finite bias or by an exponentially small signal-to-noise ratio when the wave function contains nodes or is only approximately zero. 
This implies that the tVMC cannot efficiently simulate the time evolution of physically-relevant cases such as completely polarized wave functions, states arising from digital quantum circuits or measurement processes, including the open dynamics with quantum jumps.
Subsequently, we have formalized an alternative scheme, which consists in solving an optimization problem at each time step using the infidelity distance, and we have introduced a novel stochastic estimator which makes this approach viable and scalable to large systems.
Finally, we showed that our method can solve the lack of efficient algorithms to investigate the high-entanglement phase in a protocol of non-Clifford unitary dynamics with local random measurements in 2D. 
This enables future investigation into the physics of several classes of systems, including measurement-induced phase transitions in non-trivial models above 1D and the physics of dissipative systems, all of which are currently limited by the available computational methods.
In particular, a direct application of the projected method would be the variational simulation of quantum trajectories arising from unraveling the Lindblad Master Equation.

\section{Data availability}
All the simulations have been performed using Netket 3~\cite{vicentini2022netket,netket2} with MPI and MPI4jax~\cite{Vicentini2011mpi4jax}.
The code for the p-tVMC method can be found in~\cite{SinibaldiptVMC}. 

\section{Acknowledgments}
F.V. warmly thanks A. Biella for inspirational, insightful discussions and V. Savona for arguments motivating this study. 
We gratefully acknowledge J. Nys and Z. Denis for fruitful comments and discussions.  
A.S. is supported by SEFRI under Grant No.\ MB22.00051 (NEQS - Neural Quantum Simulation).
C.G. is supported by the Swiss National Science Foundation under Grant No. 200021\_200336.
\bibliographystyle{quantum}
\bibliography{biblio.bib}

\onecolumn
\newpage
\appendix

\begin{center}
\section*{\huge Appendix}
\end{center} 

\section{Unbiased force estimate}\label{sec:unbiased_F}
An unbiased estimate for the variational forces $F_k$ can be obtained by inserting the completeness relation $\sum_{\sigma} \ket{\sigma} \bra{\sigma} = \mathds{1}$ between $\mathcal{H}$ and $\ket{\Psi_{\theta}}$, instead of between $\bra*{\partial_{\theta_k^*}\Psi_{\theta}}$ and $\mathcal{H}$ as done for the standard tVMC estimate $F_k^{\text{MC}}$, yielding: 
\begin{equation}\label{eq:F_stochastic_unbiased}
\begin{aligned}
        \Tilde{F}_{k}^{\text{MC}} &= \sum_{\sigma} \frac{\bra{{\partial_{\theta_k} \Psi_{\theta}}} \mathcal{H} \ket{\sigma} \bra{\sigma} \ket{\Psi_{\theta}}}{\bra{\Psi_{\theta}}\ket{\Psi_{\theta}}} - \sum_{\sigma, \sigma^\prime}\frac{\bra{\partial_{\theta_k} \Psi_{\theta}} \ket{\sigma} \bra{\sigma}\ket{\Psi_{\theta}}}{\bra{\Psi_{\theta}} \ket{\Psi_{\theta}}} \frac{\bra{\Psi_{\theta}}\ket*{\sigma^\prime} \bra*{\sigma^\prime} \mathcal{H} \ket{\Psi_{\theta}}}{\bra{\Psi_{\theta}} \ket{\Psi_{\theta}}} = \\
        &= \mathbb{E}_{\Pi} \bigg[
         \frac{\bra{\partial_{\theta_k}\Psi_{\theta}} \mathcal{H} \ket{\sigma}}{\bra{\Psi_{\theta}}\ket{\sigma}}\bigg] - \mathbb{E}_{\Pi} [  O^*_k(\sigma)] \mathbb{E}_{\Pi}[E_{\mathrm{loc}}(\sigma)], 
\end{aligned}
\end{equation}
where, like in the main text, $E_{\mathrm{loc}}(\sigma)$ is the local energy and $O_k(\sigma)$ are the log-derivatives of the ansatz. 
$\Tilde{F}_{k}^{\text{MC}}$ does not have a covariance form like $F_k^{\text{MC}}$, therefore it has, in general, larger statistical fluctuations when estimated using a finite number of samples. Moreover, since $O_k(\sigma)$ cannot be used to compute the first term in \cref{eq:F_stochastic_unbiased}, its computational cost is generally higher than the standard estimator. 

\section{Examples of biases in tVMC}\label{sec:bias_examples}

\subsection{One-spin system}\label{sec:singlespin}
We consider a system of $N=1$ spin $1/2$ whose state is represented by the variational ansatz $\ket{\Psi_\theta} = \alpha \ket{\downarrow} + \beta \ket{\uparrow}$ with parameters $\theta = (\alpha, \beta)$. The evolution is generated by the Hamiltonian $\mathcal{H} = \sigma_y$. 
For the choice $\alpha=1$ and $\beta=0$, we have $\ket{\Psi_{\theta}} = \ket{\downarrow}$, $\ket{\partial_{\alpha} \Psi_{\theta}} = \ket{\downarrow}$ and $\ket{\partial_{\beta}\Psi_{\theta}} = \ket{\uparrow}$.
In these conditions: 
\begin{equation}
F = \begin{pmatrix}
0\\
i
\end{pmatrix},
    \quad 
S = \begin{pmatrix}
0 & 0 \\
0 & 1 \\
\end{pmatrix}. \,\,\,
\end{equation}

However, due to the covariance form of the standard tVMC estimates $F_{k}^{\text{MC}}$ and $S_{k, k^\prime}^{\text{MC}}$ evaluated on samples including only one value of $\sigma$, we have that $\forall \, k, k^\prime$: 
\begin{equation}
    F^{\text{MC}}_k = O^*_k(\downarrow) E_{\text{loc}}(\downarrow) - O^*_k(\downarrow) E_{\text{loc}}(\downarrow) = 0, \quad
    S^{\text{MC}}_{k k^\prime} = O^*_k(\downarrow) O_{k^\prime}(\downarrow) - O^*_k(\downarrow) O_{k^\prime}(\downarrow) = 0.
\end{equation}

We verify that, instead, the stochastic estimate $\Tilde{F}^{\text{MC}}$ with the alternative estimator proposed in the previous section is unbiased, since: 
\begin{equation}
    \Tilde{F}_{\alpha}^{\text{MC}} = \frac{\bra{\partial_{\alpha}\Psi_{\theta}}\mathcal{H}\ket{\downarrow}}{\bra{\Psi_{\theta}}\ket{\downarrow}} - O_\alpha^*(\downarrow)E_{\text{loc}}(\downarrow) = 0,  \quad
      \Tilde{F}_{\beta}^{\text{MC}} = \frac{\bra{\partial_{\beta}\Psi_{\theta}}\mathcal{H}\ket{\downarrow}}{\bra{\Psi_{\theta}}\ket{\downarrow}} - O_\beta^*(\downarrow)E_{\text{loc}}(\downarrow) = i, 
\end{equation}

as $E_{\text{\text{loc}}}(\downarrow) = \mathcal{H}_{\downarrow \downarrow} \Psi_{\theta}(\downarrow) / \Psi_{\theta}(\downarrow) + \mathcal{H}_{\downarrow \uparrow} \Psi_{\theta}(\uparrow) / \Psi_{\theta}(\downarrow)  = 0$.

\subsection{Two-spin system}\label{sec:twospins}

We consider a system of $N=2$ spins $1/2$ whose state is represented by the variational ansatz $\ket{\Psi_\theta} = \alpha \ket{\downarrow \downarrow} + \beta \ket{\downarrow \uparrow} + \gamma \ket{\uparrow \downarrow} + \delta \ket{\uparrow \uparrow}$ with parameters $\theta = (\alpha, \beta, \gamma, \delta)$.
For $\theta = (1/\sqrt{2}, 0, 0, 1/\sqrt{2})$ the ansatz represents the Greenberger–Horne–Zeilinger state $\ket{\Psi_{\theta}} = \ket{\text{GHZ}} = (\ket{\uparrow \uparrow} + \ket{\downarrow \downarrow})/\sqrt{2}$. 
The variational derivatives are $\ket{\partial_{\alpha} \Psi_{\theta}} = \ket{\downarrow \downarrow}$, $\ket{\partial_{\beta} \Psi_{\theta}} = \ket{\downarrow \uparrow}$,  $\ket{\partial_{\gamma} \Psi_{\theta}} = \ket{\uparrow \downarrow}$ and $\ket{\partial_{\delta} \Psi_{\theta}} = \ket{\uparrow \uparrow}$. 
We consider the dynamics generated by the TFI Hamiltonian $\mathcal{H}_{\text{TFI}}$ with coupling $J$ and transverse field $h$. In these conditions, $F$ and $S$ differ from $F^{\text{MC}}$ and $S^{\text{MC}}$ as:
\begin{equation} \label{eq:F_S_GHZ}
F =  \begin{pmatrix}
0   \\
-h\sqrt{2}  \\
-h\sqrt{2}  \\
0   \\
\end{pmatrix},
\quad
S = \begin{pmatrix}
1/2 & 0 & 0 & -1/2\\
0 & 1 & 0 & 0\\
0 & 0 & 1 & 0\\
-1/2 & 0 & 0 & 1/2\\
\end{pmatrix}, 
\quad
F^{\text{MC}} =  \begin{pmatrix}
0   \\
0  \\
0  \\
0  
\end{pmatrix},
\quad
S^{\text{MC}} = \begin{pmatrix}
1/2 & 0 & 0 & -1/2\\
0 & 0 & 0 & 0\\
0 & 0 & 0 & 0\\
-1/2 & 0 & 0 & 1/2
\end{pmatrix}.
\end{equation}

\begin{figure}[ht!]
    \centering
    \includegraphics[scale=0.14]{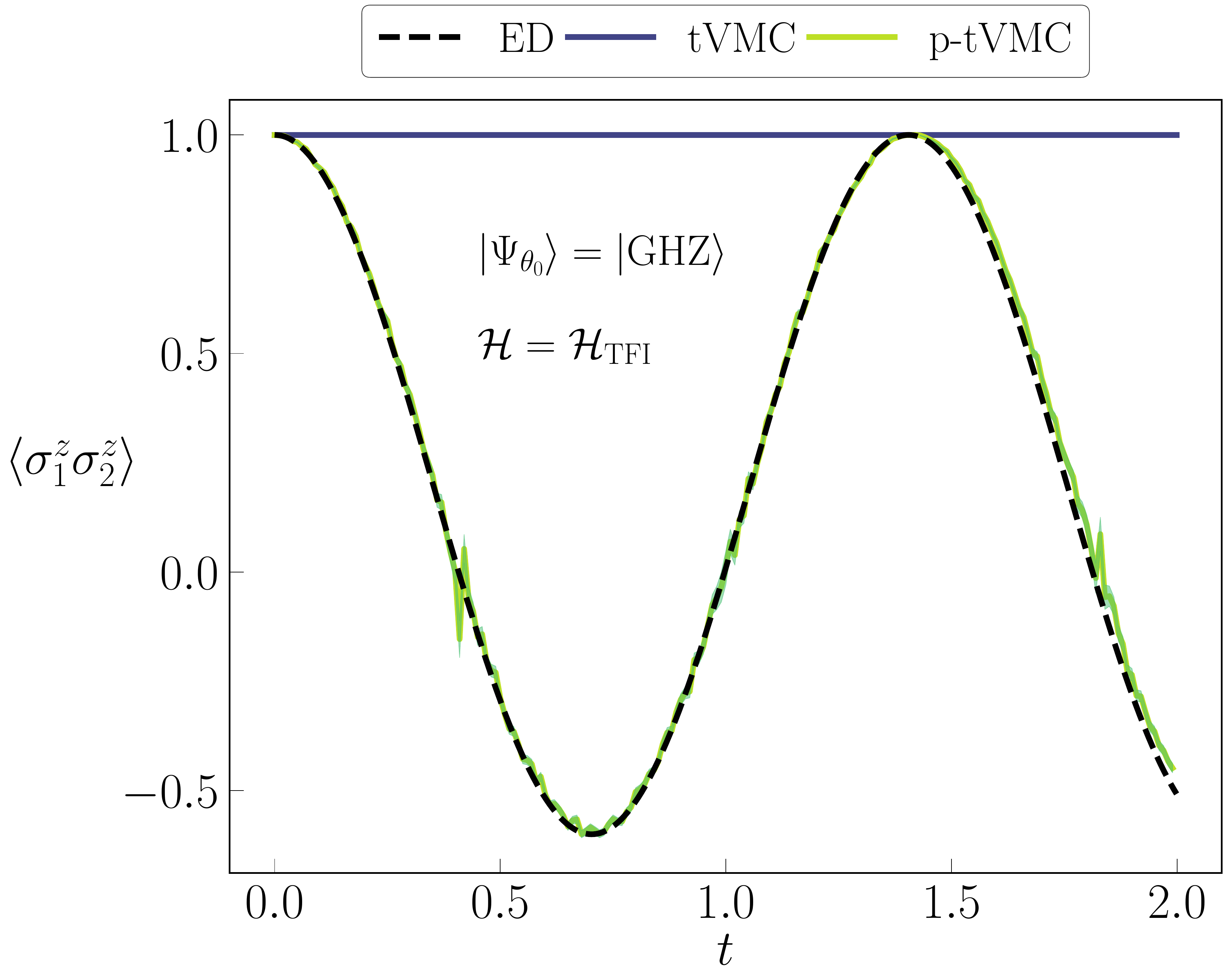}
    \caption{Dynamics of the $z$-correlator $\langle \mathcal{\sigma}_{1}^z \sigma_{2}^z \rangle$ computed with Exact Diagonalization (ED), tVMC and p-tVMC for $N=2$ spins $1/2$ evolved with $\mathcal{H}_{\text{TFI}}$ from the initial state $\ket{\text{GHZ}}$. 
    The parameters of $\mathcal{H}_{\text{TFI}}$ are $J=h=1$. The time-steps used are $\delta t= 10^{-3}$ for tVMC and $\delta t=10^{-2}$ for p-tVMC.
    $N_s = 10^3$ samples have been employed. }
    \label{fig:GHZ}
\end{figure}

One can verify that, like for the one spin case, the alternative estimator $\Tilde{F}^{\text{MC}}$ gives an unbiased estimate for the variational forces. 
As shown in \cref{eq:F_S_GHZ}, $S^{\text{MC}}$ has a lower rank than $S$, while $F^{\text{MC}}$ is identical to zero meaning that this dynamics starting from $\ket{\text{GHZ}}$ cannot be evolved with tVMC. 
On the contrary, the p-tVMC is not affected by any problem and can perform the evolution, as shown in \cref{fig:GHZ}. 

\section{Signal to noise ratios in Monte Carlo estimates}\label{sec:exploding_variance}

We consider a system of $N$ spins $1/2$ and normalized variational states $\ket{\Psi_{\varepsilon}}$ whose wave function is parameterized by a single parameter $\varepsilon$ as: 
\begin{equation}
    \Psi_{\varepsilon}(\sigma) = 
    \begin{cases}
    \sqrt{\varepsilon}, &\text{if $\sigma \neq \sigma_0$},  \\
    \sqrt{1 - (2^N - 1)\varepsilon}, &\text{if $\sigma = \sigma_0$}, 
    \end{cases}
\end{equation}
with $0<\varepsilon<1/(2^N - 1)$ and for a given $\sigma_0$. 
In the following, we prove that the signal-to-noise ratios (SNRs) of $F^{\text{MC}}_{k}$ and $S_{k k^\prime}^{\text{MC}}$ scale as $O(\sqrt{\varepsilon})$, namely they diminish indefinitely as the wave function becomes more peaked around $\sigma_0$. 
In order to ensure normalization of $\ket{\Psi_{\varepsilon}}$ over different system sizes, $\varepsilon$ goes as $1/2^N$. Therefore, the two SNRs diminish exponentially as $N$ increases, and so the number of samples needed to resolve $F_{k}^{\text{MC}}$ and $S_{k k^\prime}^{\text{MC}}$ with finite precision is exponentially large in the system size.
Instead, the SNR of the unbiased estimate $\Tilde{F}_{k}^{\text{MC}}$ of \cref{eq:F_stochastic_unbiased} is $O(1)$ in $\varepsilon$, enabling it to efficiently estimate the forces in the limit $\varepsilon \rightarrow 0$,  where $F_{k}^{\text{MC}}$ cannot be used, and independently of the system size. 

\begin{proof}
The variational log-derivative of $\Psi_{\varepsilon}(\sigma)$ is: 
\begin{equation}
    O_\varepsilon(\sigma) = 
    \begin{cases}
    \dfrac{1}{2\varepsilon}, &\text{if $\sigma \neq \sigma_0$},  \\
    -\dfrac{2^N-1}{2(1 - (2^N - 1)\varepsilon)}, &\text{if $\sigma = \sigma_0$}.
    \end{cases}
\end{equation}

We define $(1 - (2^N - 1)\varepsilon) \equiv \alpha$ for brevity and we observe that $\alpha \rightarrow 1$ when $\varepsilon \rightarrow 0$.
Both $F_k^{\text{MC}}$ and $S_{k k^\prime}^{\text{MC}}$ are of the form $\mathbb{E}_{\Pi}[\Bar{A}(\sigma) \Bar{B}(\sigma)]$, where $\Bar{A}(\sigma) = A(\sigma) - \mathbb{E}_{\Pi}[A(\sigma)]$ and $\Bar{B}(\sigma) = B(\sigma) - \mathbb{E}_{\Pi}[B(\sigma)]$ for some functions $A$ and $B$ of random variable $\sigma$ with distribution $\Pi$.
In $F_k^{\text{MC}}$ we have $A(\sigma) = O_{k}^*(\sigma)$ and $B(\sigma) = E_{\text{loc}}(\sigma)$, while in $S_{k k^\prime}^{\text{MC}}$ we have $A(\sigma) = O_{k}^*(\sigma)$ and $B(\sigma) = O_{k^\prime}^*(\sigma)$.
The variance of stochastic estimates of this form is given by: 
\begin{equation}
    \text{Var}_{\Pi}[\Bar{A}(\sigma) \Bar{B}(\sigma)] = \mathbb{E}_{\Pi}[|\Bar{A}(\sigma) \Bar{B}(\sigma)|^2] - |\mathbb{E}_{\Pi}[\Bar{A}(\sigma) \Bar{B}(\sigma)]|^2.
\end{equation}

For the local energy we have: 
\begin{align}
    E_{\text{loc}}(\sigma \neq \sigma_0) &= \sum_{\sigma^\prime \neq \sigma_0} \mathcal{H}_{\sigma \sigma^\prime} \frac{\Psi_\varepsilon(\sigma^\prime)}{\Psi_\varepsilon(\sigma)} + \mathcal{H}_{\sigma \sigma_0} \frac{\Psi_\varepsilon(\sigma_0)}{\Psi_\varepsilon(\sigma)} = \sum_{\sigma^\prime \neq \sigma_0} \mathcal{H}_{\sigma \sigma^\prime} + \mathcal{H}_{\sigma \sigma_0} \sqrt{\frac{\alpha}{\varepsilon}} \stackrel{\varepsilon \ll 1}{\approx} \mathcal{H}_{\sigma \sigma_0} \sqrt{\frac{\alpha}{\varepsilon}}, \\
    E_{\text{loc}}(\sigma_0) &= \sum_{\sigma^\prime \neq \sigma_0} \mathcal{H}_{\sigma_0 \sigma^\prime} \frac{\Psi_\varepsilon(\sigma^\prime)}{\Psi_\varepsilon(\sigma_0)} + \mathcal{H}_{\sigma_0 \sigma_0} \frac{\Psi_\varepsilon(\sigma_0)}{\Psi_\varepsilon(\sigma_0)} = \sum_{\sigma^\prime \neq \sigma_0} \mathcal{H}_{\sigma_0 \sigma^\prime} \sqrt{\frac{\varepsilon}{\alpha}}+ \mathcal{H}_{\sigma_0 \sigma_0} \stackrel{\varepsilon \ll 1}{\approx} \mathcal{H}_{\sigma_0 \sigma_0}, 
\end{align}
where the notation $\mathcal{H}_{\sigma \sigma^\prime} \equiv \bra{\sigma} \mathcal{H} \ket{\sigma^\prime}$ for any $\sigma, \sigma^\prime$ is used.
Therefore, keeping leading order terms in $1/\varepsilon$, the variances of the stochastic estimates are: 
\begin{equation}
    \text{Var}_{\Pi}[F_{k}^{\text{MC}}] \stackrel{\varepsilon \ll 1}{\approx} \frac{\alpha}{4 \varepsilon^2}  \sum_{\sigma \neq \sigma_0}|\mathcal{H}_{\sigma \sigma_0}|^2, \quad 
    \text{Var}_{\Pi}[S_{k k^\prime}^{\text{MC}}] \stackrel{\varepsilon \ll 1}{\approx} \dfrac{(2^N-1)}{16 \varepsilon^3} . 
\end{equation}

Since $F_{k}^{\text{MC}}  \approx  \sum_{\sigma \neq \sigma_0} \mathcal{H}_{\sigma \sigma_0} \sqrt{\alpha / 4 \varepsilon}$ and $S_{k k^\prime}^{\text{MC}} \approx  (2^N - 1) / 4 \varepsilon$ for $\varepsilon \ll 1 $, we have that: 
\begin{equation}
 \text{SNR}_{\Pi}[F_{k }^{\text{MC}}] \stackrel{\varepsilon \ll 1}{\approx}   \frac{|\sum_{\sigma \neq \sigma_0} \mathcal{H}_{\sigma \sigma_0}|}{\sqrt{\sum_{\sigma \neq \sigma_0} |\mathcal{H}_{\sigma \sigma_0}|^2}} \sqrt{\varepsilon} \stackrel{\varepsilon \rightarrow 0 }{\longrightarrow } 0, \quad
\text{SNR}_{\Pi}[S_{k k^\prime}^{\text{MC}}] \stackrel{\varepsilon \ll 1}{\approx} \sqrt{(2^N-1) \varepsilon} \stackrel{\varepsilon \rightarrow 0 }{\longrightarrow } 0.
\end{equation}

The two SNRs go to zero when the state becomes more peaked because when $\varepsilon \rightarrow 0$ we have that $|\Psi_\varepsilon(\sigma)|^2 \rightarrow 0$ for $\sigma \neq \sigma_0$, so these configurations are rarely sampled, but the estimators of the biases for $\sigma \neq \sigma_0$ increase as $\varepsilon$ is reduced. 
Indeed, for $\sigma \neq \sigma_0$ we have that: 
\begin{equation}
 |\partial_{\varepsilon} \Psi_\varepsilon(\sigma)|^2 = \frac{1}{4\varepsilon}  \stackrel{\varepsilon \rightarrow 0}{\longrightarrow} \infty, \quad
\partial_{\varepsilon} \Psi_\varepsilon(\sigma) \bra{\sigma} \mathcal{H} \ket{\Psi_{\varepsilon}} = \sum_{\sigma^\prime \neq \sigma_0} \frac{1}{2}\mathcal{H}_{\sigma \sigma^\prime} + \frac{1}{2}\mathcal{H}_{\sigma \sigma_0} \sqrt{\frac{\alpha}{\varepsilon}} \stackrel{\varepsilon \rightarrow 0}{\longrightarrow} \infty. 
\end{equation}

It is possible to verify that using the unbiased force estimate $\Tilde{F}_{k}^{\text{MC}}$ of \cref{eq:F_stochastic_unbiased} the SNR remains constant in the limit $\varepsilon \rightarrow 0$, making it able to efficiently compute the variational forces when the standard estimate $F_{k}^{\text{MC}}$ fails.
This is because the original sum of the first term in \cref{eq:F_stochastic_unbiased} already runs over the points where the wave function is non-zero, so to obtain the corresponding estimator it is sufficient to divide by $\bra*{\Psi_{\theta}}\ket*{\sigma}$ without excluding any point. 
Instead, to obtain the estimator of the first term in the expression of $F_{k}^{\text{MC}}$ it is necessary to divide by $|\Psi_{\theta}(\sigma)|^2$, implicitly excluding  from the Monte Carlo expression the points for which the bias $b_F$ can be non-zero.  
Considering only the contribution of the first term in $\tilde{F}_{k}^{\text{MC}}$, since this is the problematic one in the standard estimate and the second term is common to $F_{k}^{\text{MC}}$, we obtain that: 
\begin{equation}
    \text{Var}_{\Pi}[\Tilde{F}_{k}^{\text{MC}}] \stackrel{\varepsilon \ll 1}{\propto} \frac{1}{4 \varepsilon} \sum_{\sigma \neq \sigma_0} \bigg| \sum_{\sigma^\prime \neq 
    \sigma_0} \mathcal{H}_{\sigma^\prime \sigma}\bigg|^2.
\end{equation}

Therefore, since $\tilde{F}_{k}^{\text{MC}}  \approx  \sum_{\sigma \neq \sigma_0} \mathcal{H}_{\sigma \sigma_0} \sqrt{\alpha / 4 \varepsilon} = F_{k}^{\text{MC}}$ for $\varepsilon \ll 1$, we have: 
\begin{equation}
     \text{SNR}_{\Pi}[\tilde{F}_{k}^{\text{MC}}] \stackrel{\varepsilon \ll 1}{\approx}   \frac{|\sum_{\sigma \neq \sigma_0} \mathcal{H}_{\sigma \sigma_0}|}{\sqrt{\sum_{\sigma \neq \sigma_0} | \sum_{\sigma^\prime \neq 
    \sigma_0} \mathcal{H}_{\sigma^\prime \sigma}|^2}} \sqrt{\alpha} \stackrel{\varepsilon \rightarrow 0}{\longrightarrow} \text{const.} 
\end{equation}
\end{proof}

\section{Example of vanishing SNRs in tVMC} \label{sec:adiabatic}

We consider a chain of $N$ spins $1/2$, initially in the ground state of the Hamiltonian $-\sum_{i} \sigma_i^x$, which evolves according to the time-dependent Hamiltonian: 
\begin{equation}\label{eq:adiabatic_Hamiltonian}
    \mathcal{H}(t) = \gamma(t) \sum_i \sigma_i^z + (\gamma(t) - 1) \sum_i \sigma^x_i,
\end{equation} 
where $\gamma(t)$ oscillates between $0$ and $1$ with a triangular profile of period $T$, namely $\gamma(t) = t/T$ if $0<t<T$ and $\gamma(t) = 1 - (t-T)/T$ if $t>T$. 
It is known that for a sufficiently large $T$, so for an adiabatic evolution, the state at $t=T$ is going to be $\varepsilon$-close to $\bigotimes_{i=1}^N \ket{\downarrow}_i$, so an instance of the peaked states $\ket{\Psi_{\varepsilon}}$ of the previous section for some $\varepsilon$ depending on $T$.

\begin{figure}[!ht]
    \centering
    \includegraphics[scale=0.14]{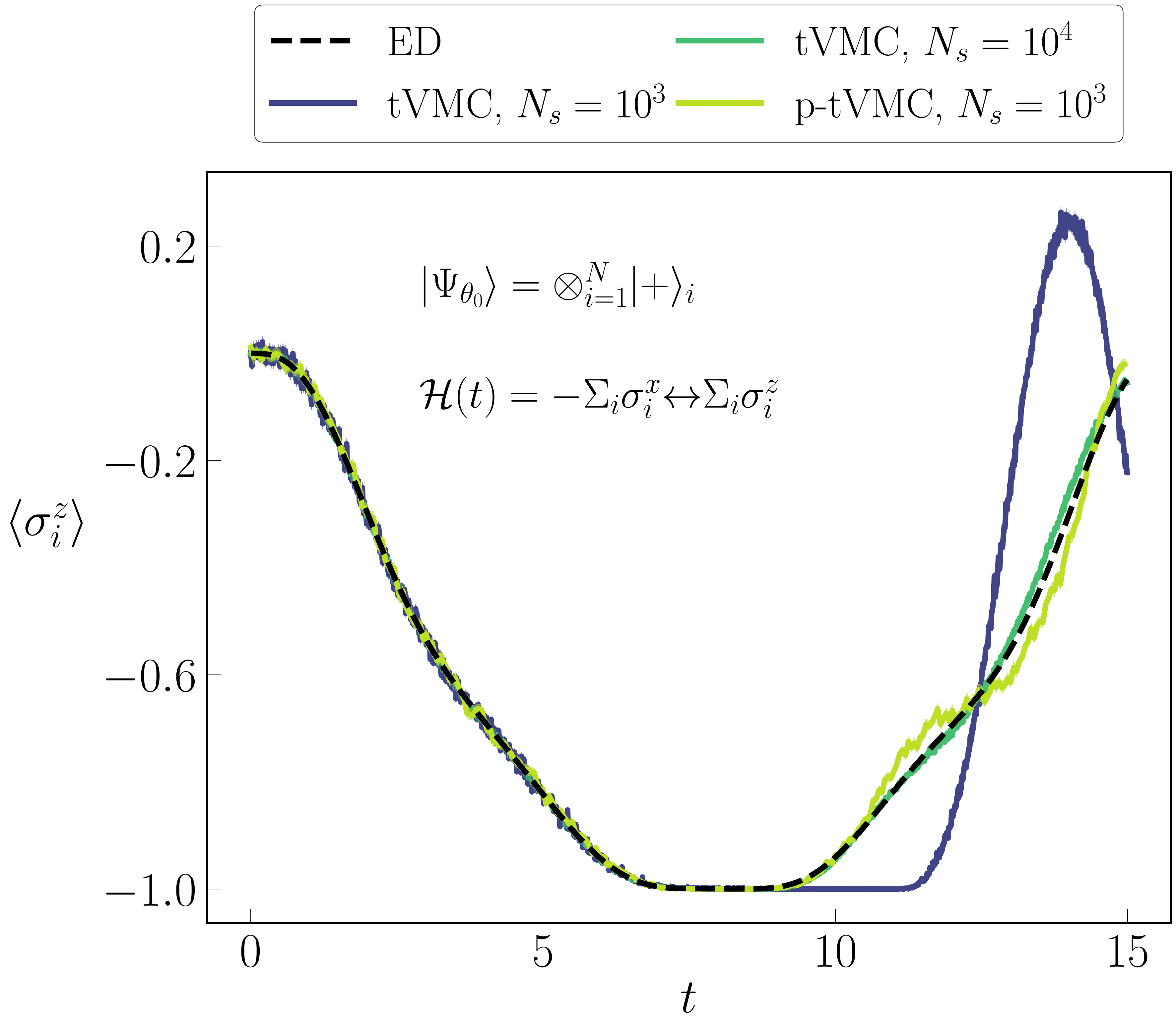}
    \caption{Dynamics of the $z$-magnetization $\langle \mathcal{\sigma}_{i}^z \rangle$ computed with Exact Diagonalization (ED), tVMC and p-tVMC for $N=10$ spins $1/2$ adiabatically evolved with $\mathcal{H}(t)$ of \cref{eq:adiabatic_Hamiltonian} with period $T=8$ from the ground state of $-\sum_{i} \sigma_i^x$. 
    The time-steps used are $\delta t= 10^{-3}$ for tVMC and $\delta t=10^{-2}$ for p-tVMC.
    The ansatz is an RBM with $\alpha = 1$. }
    \label{fig:adiabatic}
\end{figure}

While in this case the biases in the tVMC estimates are zero, in \cref{fig:adiabatic} we show that the dynamics is correctly reconstructed for $t>T$ only when choosing a sufficiently large number of $N_s=10^4$ samples (the Hilbert space in this case has size $2^{10}\approx10^3$), hinting at the exponentially small SNRs of $F^{\text{MC}}$ and $S^{\text{MC}}$. 
As before, the p-tVMC can efficiently simulate this dynamics with a fair number of $N_s=10^3$ samples, for which instead the tVMC fails. 

\section{Infidelity estimator with control variates}\label{sec:CV_estimator}
The stochastic estimate of the infidelity between the states $\ket*{\Psi_{\Tilde{\theta}}}$ and $\mathcal{U}\ket{\Psi_{\theta}}$, where $\ket*{\Psi_{\Tilde{\theta}}}$ and $\ket{\Psi_{\theta}}$ are variational states and $\mathcal{U}$ is an arbitrary transformation, is obtained by inserting completeness relations and dividing by wave function amplitudes in its definition as follows: 
\begin{equation}
    \mathcal{I} = 1 - \frac{\bra{\Psi_{\Tilde{\theta}}} \mathcal{U}\ket{\Psi_{\theta}} \bra{\Psi_{\theta}} \mathcal{U}^{\dagger} \ket*{\Psi_{\Tilde{\theta}}}}{\bra{\Psi_{\Tilde{\theta}}} \ket*{\Psi_{\Tilde{\theta}}} \bra{\Psi_{\theta}} \ket{\Psi_{\theta}}} = \sum_{\sigma, \eta} \frac{|\Psi_{\Tilde{\theta}}(\sigma)|^2}{\bra{\Psi_{\Tilde{\theta}}}\ket*{\Psi_{\Tilde{\theta}}}} \frac{|\Psi_{\theta}(\eta)|^2}{\bra{\Psi_{\theta}}\ket{\Psi_{\theta}}} \bigg[1 - \frac{\bra{\sigma} \mathcal{U} \ket{\Psi_{\theta}}}{\bra{\sigma} \ket*{\Psi_{\Tilde{\theta}}}} \frac{\bra{\eta} \mathcal{U}^{\dagger} \ket*{\Psi_{\Tilde{\theta}}}}{\bra{\eta} \ket{\Psi_{\theta}}} \bigg] = \mathbb{E}_{\chi}[\mathcal{I}_{\text{loc}}(\sigma, \eta)], 
\end{equation}
where $\mathcal{I}_{\text{loc}}(\sigma, \eta)$ is the term in brackets and $\chi(\sigma, \eta) = |\Psi_{\Tilde{\theta}}(\sigma)|^2 |\Psi_{\theta}(\eta)|^2  / \bra{\Psi_{\Tilde{\theta}}}\ket*{\Psi_{\Tilde{\theta}}} \bra{\Psi_{\theta}}\ket{\Psi_{\theta}} $. 

Similarly, the (conjugate) gradient of the infidelity can be computed stochastically as: 
\begin{equation}\label{eq:infidelity_gradient}
 \partial_{\Tilde{\theta}^{ *}_k} \mathcal{I} =\mathbb{E}_{\chi}[O^{*}_k(\sigma) \mathcal{I}_{\text{loc}}(\sigma, \eta)] - \mathbb{E}_{\chi}[O^{*}_k(\sigma)] \mathbb{E}_{\chi}[\mathcal{I}_{\text{loc}}(\sigma, \eta)], 
\end{equation}
where $O_k(\sigma) = \partial_{\Tilde{\theta}_k} \log \Psi_{\Tilde{\theta}}(\sigma)$. 
We remark that the first term in the previous expression is free from the problem present in $F^{\text{MC}}_k$ and $S^{\text{MC}}_k$ when $\ket*{\Psi_{\Tilde{\theta}}} \rightarrow \mathcal{U} \ket*{\Psi_{\theta}}$.
Indeed, in this limit we have that for $\sigma$ where $\Psi_{\Tilde{\theta}}(\sigma) = 0$ the term $\bra*{\partial_{\Tilde{\theta}_k} \Psi_{\Tilde{\theta}}} \ket*{\sigma} \bra*{\sigma} \mathcal{U} \ket{\Psi_{\theta}}$ vanishes, so the bias as in \cref{eq:bias_in_F} disappears. 
Since we consider continuous time dynamics generated by a succession of infinitesimal $\mathcal{U}$s, such that $\ket*{\Psi_{\Tilde{\theta}}} \approx \mathcal{U} \ket*{\Psi_{\theta}}$ already at the beginning of the optimizations, we can use the gradient \cref{eq:infidelity_gradient} without incurring in the issues discussed for the tVMC. 

The variance of $\mathcal{I}_{\text{loc}}$ can be directly computed from the one of $\mathcal{F}_{\text{loc}} = 1 - \mathcal{I}_{\text{loc}}$ as:
\begin{equation}
\text{Var}_{\chi}[\mathcal{I}_{\text{loc}}(\sigma, \eta)] =
         \mathbb{E}_{\chi} [|\mathcal{F}_{\text{loc}}(\sigma, \eta)|^2]  - | \mathbb{E}_{\chi} [\mathcal{F}_{\text{loc}}(\sigma, \eta)] |^2 = 1 - (1 - \mathcal{I})^2 = 2\mathcal{I} - \mathcal{I}^2.
\end{equation}

This proves that $\mathcal{I}_{\text{loc}}$ has a variance bounded above by $1$ (because $0 \leq \mathcal{I} \leq 1$), as already noted in Ref.~\cite{havlicek2022amplitude} of the main text, and that it features a zero variance principle, since $\text{Var}_{\chi}[\mathcal{I}_{\text{loc}}(\sigma, \eta)] \rightarrow 0$ when the solution is approached ($\mathcal{I} \rightarrow 0$), as the energy in VMC for ground state search. 
However, the SNR of $\mathcal{I}_{\text{loc}}$ is: 
\begin{equation}\label{eq:SNR_bare_estimator}
\text{SNR}_{\chi}[\mathcal{I}_{\text{loc}}(\sigma, \eta)] = 
\sqrt{\frac{ \mathcal{I}^2}{\text{Var}_{\chi}[\mathcal{I}_{\text{loc}}(\sigma, \eta)]}} = \sqrt{\frac{ \mathcal{I}^2}{2 \mathcal{I} - \mathcal{I}^2}}  \stackrel{\mathcal{I} \ll 1}{\sim} \sqrt{ \mathcal{I}}, 
\end{equation}
which vanishes when $\mathcal{I} \rightarrow 0$. 
This entails that when approaching the solution the number of samples $N_s$ must diverge as $1/\mathcal{I}$ to resolve the infidelity with finite accuracy.

We find that this problem of diverging sampling overhead can be eliminated by applying the Control Variates (CV) technique on $\mathcal{I}_{\text{loc}}$. 
Since $\mathcal{I}$ is real, $\Re \mathcal{I}_{\text{loc}}$ can be considered in place of $\mathcal{I}_{\text{loc}}$ in all the following.  
CV consists of adding to an estimator an additional random variable that is correlated to the estimator and for which the expectation value is known exactly, such that the fluctuations of this variable cancel out the ones of the original estimator. 
For the infidelity, we discovered that $|\mathcal{F}_{\text{loc}}|^2$ satisfies the required properties, since it is correlated to $\mathcal{I_{\text{loc}}}$ and we have $\mathbb{E}_{\chi}[|\mathcal{F}_{\text{loc}}|^2] = \mathbb{E}_{\chi}[|1-\mathcal{I}_{\text{loc}}|^2] = 1$. 
Therefore, we can define the CV infidelity estimator:
\begin{equation}
    \mathcal{I}_{\text{loc}}^{\text{CV}}  = \Re \mathcal{I}_{\text{loc}} - c (|1-\mathcal{I}_{\text{loc}}|^2 - 1), 
\end{equation}
where $c \in \mathbb{R}$. $\mathcal{I}_{\text{loc}}^{\text{CV}}$ has the same mean as $\Re \mathcal{I}_{\text{loc}}$ for any $c$, while its variance depends on $c$. 
Therefore, $c$ can be chosen such that $\text{Var}_{\chi}[\mathcal{I}_{\text{loc}}^{\text{CV}}(\sigma, \eta)]$ is minimized. 
This optimal value, say $c^{*}$, is: 
\begin{equation} \label{eq:c_star}
            c^{*} = - \frac{\text{Cov}_{\chi} [\Re \mathcal{F}_{\text{loc}}(\sigma, \eta), |\mathcal{F}_{\text{loc}}(\sigma, \eta)|^2]}{\text{Var}_{\chi}[|\mathcal{F}_{\text{loc}}(\sigma, \eta)|^2]},  
\end{equation}
where $\text{Cov}_{\chi}[A(\sigma), B(\sigma)]$ is the covariance of functions $A$ and $B$ of random variable $\sigma$ with distribution $\chi$. The corresponding minimized variance is: 
\begin{equation}
\begin{aligned}
\text{Var}_{\chi}[\mathcal{I}_{\text{loc}}^{\text{CV}}(\sigma, \eta)](c^{*}) &= \text{Var}_{\chi}[\Re \mathcal{I}_{\text{loc}}(\sigma, \eta)] \bigg(1 - \frac{{c^*}^2 \text{Var}_{\chi}[|\mathcal{F}_{\text{loc}}(\sigma, \eta)|^2]}{\text{Var}_{\chi}[\Re \mathcal{I}_{\text{loc}}(\sigma, \eta)] } \bigg), 
\end{aligned}
\end{equation}
which is  smaller than $\text{Var}_{\chi}[\Re \mathcal{I}_{\text{loc}}(\sigma, \eta)]$. 
$c^{*}$ from \cref{eq:c_star} depends on the parameters $\Tilde{\theta}$ of the variational ansatz, thus its value varies during the optimization. 
However, it is possible to prove that when $\ket*{\Psi_{\Tilde{\theta}}} \rightarrow \mathcal{U} \ket{\Psi_{\theta}}$, $c^{*}$ tends to $-1/2$, such that we can fix $c=-1/2$ when $\mathcal{I}$ is smaller than a predetermined value and avoid to re-estimate $c^*$ at each iteration. 
The limit value of $c^{*}$ can be proven analytically considering that when $\ket*{\Psi_{\Tilde{\theta}}} \rightarrow \mathcal{U} \ket{\Psi_{\theta}}$ we have $\Re \mathcal{F}_{\text{loc}} \rightarrow 1 $ and $|\mathcal{F}_{\text{loc}}|^2 \rightarrow 1 $ with same order as $\Re^2 \mathcal{F}_{\text{loc}}$. 
Therefore, computing $\lim_{\ket*{\Psi_{\Tilde{\theta}}} \rightarrow \mathcal{U} \ket{\Psi_{\theta}}} c^*$ reduces to solve $\lim_{x \rightarrow 1 } - (x^3-x)/(x^4-1) = -1/2$. 

For reasons similar to the ones previously discussed for the first term in \cref{eq:infidelity_gradient}, the estimation of the CV factor $|1- \mathcal{I}_{\text{loc}}|^2$ is not affected by a bias or a vanishing SNR as $F_{k}^{\text{MC}}$ and $S_{k k^\prime}^{\text{MC}}$ when $\ket{\Psi_{\Tilde{\theta}}} \rightarrow \mathcal{U} \ket{\Psi_{\theta}}$, and it is precisely in this limit where CV acts to keep the SNR constant. 
Therefore, $\mathcal{I}_{\text{loc}}^{\text{CV}}$ is a well-defined infidelity estimator. 

\section{Infidelity estimator with importance sampling} \label{sec:importance_sampling}
The estimator $\Re \mathcal{I}_{\text{loc}}(\sigma, \eta)$ is unbounded because it contains ratios of wave function amplitudes. 
Therefore, it may diverge if $\bra{\sigma} \mathcal{U} \ket{\Psi_{\theta}}$ and $\bra{\sigma} \ket*{\Psi_{\Tilde{\theta}}}$ or $\bra{\eta} \mathcal{U}^{\dagger} \ket*{\Psi_{\Tilde{\theta}}}$ and $\bra{\eta} \ket*{\Psi_{\theta}}$ differ of orders of magnitudes. 
This is a severe problem when evolving the variational  state after local measurements, since it may happen that on some configurations the state $\ket{\Psi_{\theta}}$ is close to zero but $\mathcal{U}^{\dagger} \ket*{\Psi_{\Tilde{\theta}}}$ is not. Therefore, despite these configurations are rarely sampled, the estimator $\mathcal{I}_{\text{loc}}(\sigma, \eta)$ on them can be very large, significantly skewing the statistics. 
A way to include these outliers in the sampling is to perform importance sampling (see Ref.~\cite{rubinstein2016simulation} of the main text), namely by rewriting: 
\begin{equation}
    \mathcal{I} = \mathbb{E}_{\chi} [\mathcal{I}_{\text{loc}}(\sigma, \eta)] = \mathbb{E}_{\chi^{\prime}} \bigg[\mathcal{I}_{\text{loc}}(\sigma, \eta)\frac{\chi(\sigma, \eta)}{\chi^{\prime}(\sigma, \eta)}\bigg], 
\end{equation}
for a given distribution $\chi^{\prime}$. This latter can be chosen such that it minimizes the variance of the estimator, leading to the expression: 
\begin{equation}\label{eq:importance_sampling}
    \mathcal{I} = \mathbb{E}_{\chi^{\prime}}[ \mathcal{I}_{\text{loc}}(\sigma, \eta) w(\sigma, \eta) ], 
    \quad
    \chi^{\prime}(\sigma, \eta) = \frac{| \mathcal{I}_{\text{loc}}(\sigma, \eta)| \chi(\sigma, \eta)}{\mathbb{E}_{\chi}[|\mathcal{I}_{\text{loc}}(\sigma, \eta)|]}, 
\end{equation}
where $w(\sigma, \eta) = \mathbb{E}_{\chi}[|\mathcal{I}_{\text{loc}}(\sigma, \eta)|] /  |\mathcal{I}_{\text{loc}}(\sigma, \eta)|$. 
Using $\mathcal{I}_{\text{loc}}^{\text{CV}}$ in place of $\mathcal{I}_{\text{loc}}$, importance sampling can be combined with control variates obtaining the estimator: 
\begin{equation}
    \mathcal{I} = \mathbb{E}_{\chi^{\prime}}[\mathcal{I}_{\text{loc}}^{\text{CV}}(\sigma, \eta) w(\sigma, \eta)]. 
\end{equation}

The cost of importance sampling is almost $n$ times larger than the cost of standard sampling, where $n$ is the number of spins on which $\mathcal{U}$ acts non-trivially.

\section{Exact application of a diagonal propagator}\label{sec:exact_application}
The exact solution $\Tilde{\theta}$ of the general optimization problem in the p-tVMC must satisfy: 
\begin{equation}\label{eq:propagator_equation}
        \bra{\sigma}\ket*{\Psi_{\Tilde{\theta}}}=C\bra{\sigma} \mathcal{U} \ket{\Psi_{\theta}}, 
\end{equation}
for all configurations $\sigma$ and any constant $C$ equal for all $\sigma$. 
For certain $\mathcal{U}$ and variational states, \cref{eq:propagator_equation} can be solved exactly. 
In particular, if $\mathcal{U}$ is diagonal in the basis $\{\ket{\sigma}\}$ and if it is possible to write $\bra{\sigma}\ket*{\Psi_{\Tilde{\theta}}} = \bra{\sigma}\ket{\Psi_{\theta}} \bra{\sigma}\ket{\Psi_{\delta \theta}}$ $\forall \sigma$ with some update $\delta \theta$, \cref{eq:propagator_equation} is reduced to:
\begin{equation}\label{eq:propagation_equation_diagonal}
\bra{\sigma} \ket{\Psi_{\delta \theta}} = C U_\sigma^{*}, 
\end{equation}
where $U_\sigma$ is the eigenvalue of $\mathcal{U}$ for the eigenstate $\ket{\sigma}$. 
In the following, we consider spin systems for simplicity.  
For the RBM, \cref{eq:propagation_equation_diagonal} admits a solution when $\mathcal{U} = R_{i}^{z}(\phi_{z}) = e^{-i\phi_z \sigma_i^z}$ consisting in an update of the visible bias only. 
If $\mathcal{U} = R_{ij}^{zz}(\phi_{zz}) = e^{-i\phi_{zz} \sigma_i^z \sigma_j^z}$, instead, a simple parameter update is not sufficient. The transformation can be exactly implemented by adding a hidden unit (see Refs.~\cite{jonsson2018neural,medvidovic2021classical} of the main text). 
However, in general, starting with an arbitrary ansatz $\ket{\Psi_{\theta}}$, it is possible to add two variational terms such that both $R_{i}^{z}$ and $R_{ij}^{zz}$ can be exactly simulated with a parameter change. 
Indeed, $\ket{\Psi_{\theta}}$ can be modified into a new ansatz $|\Psi_{J^{(1)}_i, J^{(2)}_{ij}, \theta}\rangle$ with two additional parameters $\{J^{(1)}_i, J^{(2)}_{ij}\}$ which is defined as: 
\begin{equation}\label{eq:extended_ansatz}
    |\Psi_{J^{(1)}_i, J^{(2)}_{ij}, \theta}\rangle = e^{- i J^{(1)}_i \sigma^z_i } e^{- i J^{(2)}_{ij} \sigma^z_i \sigma^z_j}  \ket{\Psi_{\theta}}. 
\end{equation}

The two exponential terms factorize as required to have \cref{eq:propagation_equation_diagonal}, such that the application of $R_{i}^{z}(\phi_z)$ translates into the updates $\delta J^{(1)}_i = \phi_z$, $\delta J^{(2)}_{ij} = 0$ and $\delta \theta = 0$, while for $R_{ij}^{zz}(\phi_{zz})$ the changes $\delta J^{(1)}_i = 0$, $\delta J^{(2)}_{ij} = \phi_{zz}$ and $\delta \theta = 0$ are required.  
Adding many two-site terms in the ansatz, it is possible to simulate the dynamics of the diagonal part of the TFI Hamiltonian $\mathcal{H}_{\text{TFI}}$ exactly.

\section{Stochastic estimator for the Rényi-2 entropy}\label{sec:Reniy2}
An estimator for the Rényi-2 entanglement entropy $S_2 = - \log_2 \Tr \rho_A^2$ can be obtained by explicitly writing the definition of the purity of the reduced density matrix: 
\begin{equation}
    \Tr \rho_A^2 = \Tr[ \bigg(\Tr_{B}  \frac{\ket{\Psi_{\theta}} \bra{\Psi_{\theta}}}{\bra{\Psi_{\theta}} \ket{\Psi_{\theta}}}\bigg)^2] 
    = \sum_{\sigma^\prime, \eta, \eta^{\prime}} \frac{\bra{\sigma^\prime, \eta} \ket{\Psi_{\theta}} \bra{\Psi_{\theta}} \ket{\eta} \bra{\eta^\prime}\ket{\Psi_{\theta}} \bra{\Psi_{\theta}} \ket{\eta^\prime, \sigma^\prime}}{\bra{\Psi_{\theta}} \ket{\Psi_{\theta}} \bra{\Psi_{\theta}} \ket{\Psi_{\theta}}}, 
\end{equation}
where $\Tr_{B}$ indicates the partial trace over $B$, $\sigma^\prime \in A$ and $\eta, \eta^\prime \in B$. 
Now, inserting a further completeness relation in $A$ and dividing by wave function amplitudes one obtains: 
\begin{equation}
\begin{aligned}
    \Tr \rho_A^2 &= \sum_{\substack{\sigma, \sigma^{\prime} \\ \eta, \eta^{\prime}}}  \frac{|\Psi_{\theta}(\sigma, \eta)|^2}{\bra{\Psi_{\theta}} \ket{\Psi_{\theta}}} \frac{|\Psi_{\theta}(\sigma^\prime, \eta^\prime)|^2}{\bra{\Psi_{\theta}} \ket{\Psi_{\theta}}} \frac{\Psi_{\theta}(\sigma^{\prime}, \eta) \Psi_{\theta}(\sigma, \eta^{\prime})}{\Psi_{\theta}(\sigma, \eta) \Psi_{\theta}(\sigma^{\prime}, \eta^{\prime})} = \mathbb{E}_{\substack{\Pi(\sigma, \eta) \\ \hspace{0.2cm}\Pi(\sigma^\prime, \eta^\prime)}} \bigg[ \frac{\Psi_{\theta}(\sigma^{\prime}, \eta) \Psi_{\theta}(\sigma, \eta^{\prime})}{\Psi_{\theta}(\sigma, \eta) \Psi_{\theta}(\sigma^{\prime}, \eta^{\prime})}\bigg].
\end{aligned}
\end{equation}

\end{document}